\documentclass[%
twocolumn,
aps,
prx,
superscriptaddress,
showkeys,
longbibliography,
]{revtex4-2}

\usepackage{packages}

\begin{document}
\title{The parametric instability landscape of coupled Kerr parametric oscillators}
\author{Orjan Ameye}
\affiliation{Department of Physics, University of Konstanz, 78464 Konstanz, Germany}
\author{Alexander Eichler}
\affiliation{Laboratory for Solid State Physics, ETH Z\"{u}rich, CH-8093 Z\"urich, Switzerland.}
\author{Oded Zilberberg}
\affiliation{Department of Physics, University of Konstanz, 78464 Konstanz, Germany}
\date{\today}

\begin{abstract}
  Networks of coupled Kerr parametric oscillators (KPOs) hold promise as for the realization of neuromorphic and quantum computation. Yet, their rich bifurcation structure remains largely not understood. Here, we employ secular perturbation theory to map the stability regions of these networks, and identify the regime where the system can be mapped to an Ising model. Starting with two coupled KPOs, we show how the bifurcations arise from the competition between the global parametric drive and linear coupling between the KPOs. We then extend this framework to larger networks with all-to-all equal coupling, deriving analytical expressions for the full cascade of bifurcation transitions. In the thermodynamic limit, we find that these transitions become uniformly spaced, leading to a highly regular structure. Our results reveal the precise bounds under which KPO networks have an Ising-like solution space, and thus provides crucial guidance for their experimental implementation.
\end{abstract}

\maketitle

%%%%%%%%%%%%%%%%%%%%%%%%%%%%%%%%%%%%%%%%%%%%%%%%%%%%%%%%%%%%%%%%%%%%%%
%                       Introduction
%%%%%%%%%%%%%%%%%%%%%%%%%%%%%%%%%%%%%%%%%%%%%%%%%%%%%%%%%%%%%%%%%%%%%%
\section{Introduction}
The Kerr Parametric Oscillator (KPO) stands out as a ubiquitous out-of-equilibrium system in the field of quantum mechanics and nonlinear dynamics~\cite{DykmanBook,eichler2023classical}. This system, driven by a time-dependent harmonic potential term, has emerged as a versatile platform with applications ranging from quantum information processing~\cite{Grimm_2019,kanaoQuantum2022}, metrology~\cite{GuoQuantum2024, dicandiaCriticalParametricQuantum2023}, and ultra-sensitive detection~\cite{Mahboob_2008,karabalin_2011,eichlerParametric2018}, to neuromorphic computing~\cite{DykmanInteraction2018,gotoCombinatorial2019}. In quantum information, KPOs serve as qubits and quantum gates, while in sensing applications, they enable precision measurements through parametric amplification. Their capability to form networks with complex dynamics has made them particularly attractive for neuromorphic architectures, where they can implement both computational and memory functions.

When the KPO is driven above its parametric driving threshold, the system is forced across a spontaneous period-doubling $\mathbb{Z}_2$ symmetry-breaking phase transition~\cite{ParametricLeuch2016,SorienteDistinctive2021,DykmanInteraction2018}, resulting in a ``phase state'' oscillation solution at half the drive frequency. Intriguingly, a KPO features two such phase states with equal amplitude, but phases separated by an angle $\pi$. These states can be mapped to the two polarization states of an Ising spin, ``up'' and ``down''. This analogy has sparked interest in KPO networks as a tool to solve NP-hard problems, such as finding the ground state of an Ising Hamiltonian, where different classes of ``KPO Ising machines'' with bilinear or dissipative coupling emerged~\cite{Mahboob_2016,Inagaki_2016_Science,Nigg_2017,Puri_2017,DykmanInteraction2018,Bello_2019,okawachi2020demonstration,HeugelIsing2022,Margiani_2023,alvarez2023biased,han2023controlled}. The exploration of coupled KPOs thus opens up new possibilities in the realm of computation and optimization, making it a captivating subject of research.

%\remark{3) twist in the plot, e.g., open challenge / new approach}
A key challenge for using KPO networks as Ising machines is to understand the plethora of phases in the driven many-body system, and how they map to a desired Ising Hamiltonian. Previous studies on networks with bilinear coupling have focused mainly on the case of weak coupling coefficients, where the bare modes of the system can be readily mapped to Ising spin configurations~\cite{DykmanInteraction2018}. In contrast, a strong bilinear coupling between KPOs was found to give rise to additional phases beyond the Ising model~\cite{HeugelIsing2022,Margiani3PO2025}. In particular, it became clear that the number of maximum stable configurations of an $N$-node KPO network deviates from the expected solution space of the $2^N$ Ising states. This discrepancy arises on the one hand because the resonator network forms normal modes that are separated in frequency~\cite{heugel_classical_2019}, and on the other hand because the interplay between nonlinearity and resonator coupling generates a host of additional states with mixed symmetry~\cite{HeugelIsing2022}. Even for only two KPOs, the combination of these effects results in a complex phase diagram that so far is only partially understood. As such, a deep analytical understanding of the phases associated with two coupled KPOs is crucial for predicting the behavior of larger networks, and eventually for understanding the rules governing large-scale Ising machines.

%\remark{ 4) a new hope.}
%In the regime of strong coupling, exploring the system through the lens of the normal mode basis emerges as a natural approach. This effective description accentuates the intricate interplay between the nonlinearity and node-to-node coupling within the network, resulting in nonlinear interactions between modes. These interactions between phase states critically determine the stability and nature of configurations within the network. Recently, it was proven that close to the parametric threshold the maximum number of configuration in a KPO network of size $N$ is capped by $3^N$~\cite{BreidingAlgebraic2022,borovik2023khovanskii}. This scaling property offers a means to represent the state space of the network as a tensor product of spin-1 configurations. Together with the normal mode basis, this state space representation becomes instrumental in categorizing and understanding these configurations. Moreover, this basis enables the investigation of the stability of these states under the influence of nonlinear coupling, shedding light on their behavior within the network.

%\remark{5) Summarize the results}
In this work, we present a comprehensive analysis of the phase diagrams of networks of coupled bilinear Kerr parametric oscillators (KPOs). We use secular perturbation theory to understand the stability of the different states in the system, focusing on the interplay between parametric driving and coupling. By examining the case of two coupled KPOs, we derive the conditions under which different stationary states emerge and the physical mechanisms behind them. We extend our analysis to larger networks with all-to-all equally coupling, finding analytical expressions for the bifurcation lines that demarcate the parameter regime where the network can effectively function as an Ising machine. Our findings offer a detailed understanding of the dynamics and solution space of KPO networks, paving the way for their practical implementation in neuromorphic computing applications.

The paper is structured as follows: In Sec.~\ref{sec: section 2}, we provide a comprehensive introduction to linearly coupled KPO networks, where we highlight how cross-Kerr coupling [cf.~Eq.~\eqref{eq: nonlinear terms}] due to strong coupling leads to a rich bifurcation structure. We proceed in Sec.~\ref{sec: 2 id KPO}, by analyzing the stability phase diagram of two coupled KPOs, deriving analytical expressions in Eq.~\eqref{eq: bif lines symmetric state} for the bifurcation lines that determine when different states become stable. Crucially, the stability of the states is perturbed by the coupling between the KPOs. 
In Sec.~\ref{sec: N id J}, we extend this result to networks of N identical all-to-all coupled KPOs, thus revealing a cascade of bifurcation transitions, cf.~Eq.~\eqref{eq: bif shift}. We find that the bifurcation lines~\eqref{eq: bif line limit} and \eqref{eq: bif line bare modes} define lower bounds on where the system has $2^N$ possible stationary states, thus defining the parameter regime where the system can be mapped to an Ising network.
We conclude with a summary and outlook in Sec.~\ref{sec: conclusion}.

%%%%%%%%%%%%%%%%%%%%%%%%%%%%%%%%%%%%%%%%%%%%%%%%%%%%%%%%%%%%%%%%%%%%%%
%                        Main Body
%%%%%%%%%%%%%%%%%%%%%%%%%%%%%%%%%%%%%%%%%%%%%%%%%%%%%%%%%%%%%%%%%%%%%%

\section{Coupled Kerr Parametric oscillators} \label{sec: section 2}
We analyze a network of $N$ coupled KPOs, see Fig.~\ref{fig: Fig_1}(a). Its classical equations of motion (EOMs) read
%
% \begingroup  \setmuskip{\medmuskip}{0mu}
\begin{equation}\label{eq: eom KPO}
	\ddot{x}_i + \omega_{i}^2(t) x_i +\gamma_i\dot{x}_i+ \alpha_i x_i^3  - \sum_{j \neq i} J_{i j} x_j=0\,,
\end{equation}
% \endgroup
%
where the $i^{\rm th}$ EOM describes an KPO by its displacement $x_i$, modulated bare frequency $\omega_{i}(t)=\omega_i(1-\tilde{\lambda_i}\cos(2\omega t))$, and Kerr nonlinearity $\alpha_i$. We work in units such that the mass of each KPO is one, $m_i=1$.
The dots indicate derivatives with respect to time $t$.
All KPOs are parametrically driven at the same angular frequency $2\omega$ with parametric pump strengths $\tilde{\lambda}_i\equiv\lambda_i/\omega_i^2\equiv \lambda/\omega^2_i$.
The network is formed by linear coupling between the KPOs with coefficients $J_{i j}$.
To account for dissipation in the system, we introduced linear damping terms with coefficients $\gamma_i$.

For completeness, we present our results also using equivalent quantum notation by quantizing the phase space coordinates $x_i$ and $p_i\equiv \dot{x}_i$ in Eq.~\eqref{eq: eom KPO}. Namely, we employ raising and lowering operators $\hat{x}_i \equiv \sqrt{\hbar/2\omega}\,(\hat{a}_i^\dagger+\hat{a}_i)$ and $\hat{p}_i \equiv \sqrt{\hbar\omega/2}\,(\hat{a}_i^\dagger-\hat{a}_i)$. The resulting Hamiltonian (for $\gamma_i=0$) reads
% \begingroup  \setmuskip{\medmuskip}{0mu}
\begin{multline} \label{eq: non-RWA Qham}
	\frac{H}{\hbar} =\sum_i \frac{\omega_{i}^2(t)+\omega^2}{2\omega} \hat{a}_i^\dagger \hat{a}_i + \frac{\omega_{i}^2(t)-\omega^2}{4\omega} (\hat{a}_i^\dagger \hat{a}_i^\dagger+\hat{a}_i\hat{a}_i)\\
	+ \frac{U_i}{12}(\hat{a}_i^\dagger + \hat{a}_i)^4 +\frac{1}{2}\sum_{j\neq i} T_{i j}(\hat{a}_i^\dagger +\hat{a}_i)(\hat{a}_j^\dagger +\hat{a}_j)
\end{multline}
% \endgroup
% \begin{multline}
%   \frac{H}{\hbar} = \frac{\omega_{i}^2(t)+\omega^2}{2\omega} \hat{a}^\dagger \hat{a} + \frac{\omega_{i}^2(t)-\omega^2}{4\omega} (\hat{a}^\dagger \hat{a}^\dagger+\hat{a}\hat{a})+ \frac{U}{12}(\hat{a}^\dagger + \hat{a})^4,
% \end{multline}
where $U_i =3\alpha_i\hbar/4 \omega^2$ and $T_{i j}=J_{i j}/2\omega$. Here, we quantized the system in ladder operators at the frequency of the drive $\omega$, anticipating the frequency of the response of the oscillators~\cite{Kosata_Fixing_2022,seibold2024floquetexpansioncountingpump}.

%%%%%%%%%%%%%%%%%%%%%%%%%%%%%%%%%%%%%%%%%%%%%%%%%%%%%%%%%%%%%%%%%%%%%%%

\subsection{The linear regime}
\label{sec:stability linear}
We first recall the linear limit ($\alpha_i=0$) of a single decoupled mode ($J_{ij}=0$) in the resonator network.
The EOM of the individual oscillator [cf.~Eq.~\eqref{eq: eom KPO}] takes the form of a damped Mathieu's equation~\cite{kovacic_mathieus_2018, eichler2023classical}.
The equation is solvable using Floquet theory, where the solutions evolve according to two Floquet characteristic exponents denoted as $\mu_{1,2}\in\mathbb{C}$~\cite{richards_Analysis_1983}.
The linear stability of the mode under the parametric drive hinges on the competition between damping and parametric amplification, which determines the sign of the real parts of $\mu_{1,2}$, also known as the Lyapunov exponents~\cite{datserisNonlinear2022a}. Specifically, by setting
\begin{equation} \label{eq: instability_condition_lyapunov}
	\operatorname{Re}\left\{\mu_{1,2}\right\}=0\,,
\end{equation}
we obtain the boundary that defines the parametric instability lobe, commonly referred to as ``Arnold tongue'', beyond which the linear mode's oscillation becomes unbounded~\cite{richards_Analysis_1983,zerbe1995brownian}.
In other words, outside the lobe, we have $\operatorname{Re}\left\{\mu_{1,2}\right\}<0$, and the mode's motion is stable.
In contrast, within the Arnold tongue, we have $\operatorname{Re}\left\{\mu_{1}\right\}>0$ or $\operatorname{Re}\left\{\mu_{2}\right\}>0$, and the mode experiences \textit{parametric resonance}, leading to an unbounded amplitude growth in the linear case.
These lobes are centered around $\omega \approx \omega_i/n$ for $n\in\mathbb{N}$. Our focus here is on the primary lobe associated with $n\approx1$, where the effects of parametric driving are most pronounced.
\begin{figure}[tbp]
	\centering
	\includegraphics[width=1.0\linewidth]{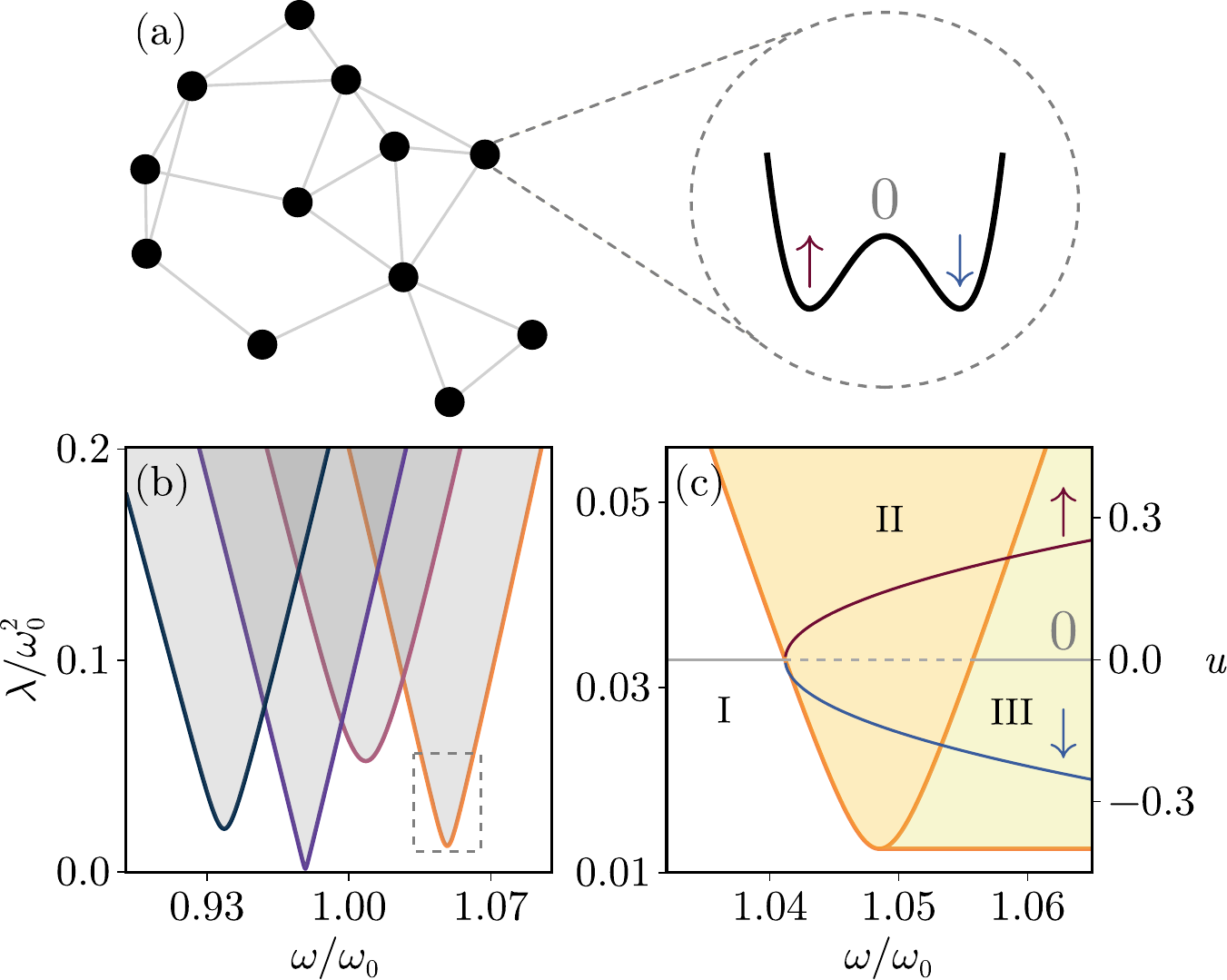}
	\caption{
		Characterization of a KPO network [cf.~Eqs.~\eqref{eq: eom KPO} and~\eqref{eq: non-RWA Qham}]. (a)~Sketch of a coupled KPO network, where the motion of each mode is effectively described via a rotating quasienergy potential which has up to three possible stable states. The latter form an effective spin-1 subspace $\{\n,\s, 0\}$, see Sec.~\ref{sec: state space}. (b)~The stability diagram around the first instability lobe for different linear parametric resonators, e.g., modes of a decoupled KPO network, or linear normal mode of the coupled system, as a function of the parametric pump amplitude $\lambda$ and driving frequency $\omega$, see Sec.~\ref{sec:stability linear}. The gray regions indicate where the $k^{\rm th}$ linear mode becomes parametrically unstable.
		(c)~The I (white), II (orange), and III (yellow) regions denote parameter regimes, where only the zero amplitude state, only the phase states, or both zero amplitude and phase states are stable, respectively [see Sec.~\eqref{sec: Beyond parametric instability}]. The quadrature $u_k$ of the stationary-state solutions as a function of $\omega$ for a cut at $\lambda=0.03$ is superimposed onto the phase diagram.
	}
	\label{fig: Fig_1}
\end{figure}

In Fig.~\ref{fig: Fig_1}(b), we show the stability diagram of linear modes in a decoupled resonator network. Note that, for each isolated mode, the parametric resonance condition occurs at a different detuning due to the different bare frequencies $\omega_i$.
Due to the different $\gamma_i$, each mode has a different \textit{parametric threshold}, namely, a parametric drive strength beyond which the $i^{\mathrm{th}}$ mode can become unstable at its primary lobe, $\lambda^{(i)}_\mathrm{th}=2\gamma_i\omega_i$.
If $\lambda_i<\lambda^{(i)}_\mathrm{th}$, no parametric excitation of the $i-th$ mode is possible over the whole frequency range.
In the case of an undamped system ($\gamma_i=0$), the parametric lobe extends to $\lambda^{(i)}_\mathrm{th}=0$, touching the frequency axis at $\omega=\omega_i$, see Fig.~\ref{fig: Fig_1}(b).

Moving to the coupled linear resonator network, the EOMs can be written in matrix formulation as $\ddot{\vb{x}}+\vb{\Gamma}\dot{\vb{x}}+\vb{M} \vb{x}-\vb{\Lambda}\cos(2\omega t)\vb{x}=0$, where $\vb{x}$ is a vector with elements $x_i$, $\vb{\Gamma}$ and $\vb{\Lambda}$ are diagonal matrices of $\gamma_i$ and $\lambda_i$, respectively, and $\vb{M}$ is a matrix containing the eigenfrequencies $\omega_i$ and off-diagonal couplings $J_{ij}$ [cf.~Eq.~\eqref{eq: eom KPO}]. When the linear resonators are strongly coupled, the problem is better characterized using the normal mode coordinates $\tilde{\vb{x}}=\vb{U}\vb{x}$, defined by the eigenvalue problem $\vb{M}\tilde{\vb{x}}=\nu\tilde{\vb{x}}$. The resulting eigenfrequencies $\nu$ only depend on the bare frequencies $\omega_i$ and couplings $J_{ij}$.
The orthogonal matrix $\vb{U}$ rotates the bare basis $x_i$ such that the linear system is decoupled~\cite{heugel_classical_2019}. This leads to a new set of $N$ EOMs for the normal modes
\begin{align}
	\ddot{x}_k +\nu_k^2x_k + \sum_{l}(\tilde{\Gamma}_{lk}\dot{x}_l-\tilde{\Lambda}_{lk} \cos(2\omega t)x_l)
	=0\,,
	\label{eq: normalEOMs}
\end{align}
where the matrices with tilde are written in the rotated basis , $\tilde{\vb{B}}=\vb{U}^T\vb{B}\vb{U}$, and each normal mode moves at its own eigenfrequency $\nu_k$. To avoid confusion, we iterate the bare mode basis with index $i$, and the normal mode basis with index $k$.

Inhomogeneous damping coefficients $\gamma_i$ lead to dissipative coupling between the normal modes. Such coupling vanishes for homogeneous damping constants, $\gamma_i=\gamma$, i.e., $\tilde\Gamma_{lk}=\gamma\delta_{lk}$ becomes diagonal~\footnote{When $\gamma_i=\gamma$, we have $\tilde{\vb{\Gamma}} = \vb{U}^T\vb{\Gamma}\vb{U} = \gamma\vb{U}^T\vb{I}\vb{U} = \gamma\vb{I}$.}. Similarly, inhomogeneous parametric drive amplitudes couple the normal modes, but are diagonal $\tilde\Lambda_{lk}=\lambda \delta_{lk}$ under a global driving amplitude, $\lambda_i=\lambda$. Neglecting such inhomogeneous coupling terms, we once more obtain a system of decoupled parametric oscillators with a stability diagram akin to Fig.~\ref{fig: Fig_1}(b), but with lobes centered around the eigenfrequencies $\nu_k$. In other words, the linear instability lines are plainly shifted due to the coupling coefficients $J_{ij}$.
% In Appendix Sec.~\ref{sm: inhomogeneous}, we explore the impact of an inhomogeneous system.
Crucially, until now, we disregarded the nonlinear terms which will couple the normal modes to one another through the transformed nonlinear terms $\sim x^3$. Their impact on the system is the main focus of this work.

%%%%%%%%%%%%%%%%%%%%%%%%%%%%%%%%%%%%%%%%%%%%%%%%%%%%%%%%%%%%%%%%%%%%%%

\subsection{Beyond parametric instability}
\label{sec: Beyond parametric instability}
The unbounded growth inside the parametric instability lobes is prevented by the nonlinearities $\alpha_i\neq 0$.
Already in the decoupled network case, $J_{ij}=0$, we obtain a rich response for a single mode, see Fig.~\ref{fig: Fig_1}(c).
This driven nonlinear system can be treated perturbatively using Floquet expansions such as the Krylov-Bogoliubov (KB) averaging method~\cite{fedorchenkoMethod1957,burshteinHamiltonian1962,mitropolskyAveraging1967,holmesSecond1981,seibold2024floquetexpansioncountingpump}.
In this approach, we transition to a rotating frame with the quadratures $u_i$ and $v_i$ defined as $x_i\equiv u_i\cos(\omega t)+v_i\sin(\omega t)$.
By averaging these quadratures over a single period $T=2\pi/\omega$, we extract the (slow) stroboscopic motion of the system and obtain slow-flow equations of the form
\begin{align} \label{eq: slow-flow equations1}
	\dot{u}_i & =-\frac{\gamma u_i}{2}-\left(\frac{3 \alpha_i}{8 \omega} A_i^2+\frac{\omega_i^2-\omega^2}{2 \omega}+\frac{\lambda}{4 \omega}\right) v_i\,,                                  \\
	\dot{v}_i & =-\frac{\gamma v_i}{2}+\left(\frac{3 \alpha_i}{8 \omega} A_i^2+\frac{\omega_i^2-\omega^2}{2 \omega}-\frac{\lambda}{4 \omega}\right) u_i\,, \label{eq: slow-flow equations2}
\end{align}
where $A_i^2=u_i^2+v_i^2$. % and the quadratures contain the lowest-order remaining time dependence of the oscillators.
The new effective description is valid when the dimensionless quantities $\abs{\omega^2-\omega_{i}^2}/\omega^2$, $\lambda/\omega^2$, $\gamma / \omega$, $J_{i j} / \omega^2$, and $\left(\alpha_i / \omega^2\right) x_i^2$ are of order $\epsilon$, where $0<\epsilon \ll 1$~\footnote{The KB method allows for higher order EOMs in $\epsilon$, however, we restrict ourselves to the first order dynamics.}.

Applying the same formalism to the quantum Hamiltonian in Eq.~\eqref{eq: non-RWA Qham}, one obtains under the same approximations a driven Bose-Hubbard-like Hamiltonian~\cite{EckardtHighFrequency2015}:
\begin{multline} \label{eq: RWA Qham}
	\frac{H_\mathrm{RWA}}{\hbar} =\sum_i \Delta_i \hat{a}_i^\dagger \hat{a}_i + \frac{G_i}{2} (\hat{a}_i^\dagger \hat{a}_i^\dagger+\hat{a}_i\hat{a}_i)\\
	+ \frac{U_i}{2}\hat{a}_i^\dagger\hat{a}_i^\dagger \hat{a}_i \hat{a}_i +\frac{1}{2}\sum_{j\neq i} T_{i j}(\hat{a}_i^\dagger\hat{a}_j + \hat{a}
	_i\hat{a}_j^\dagger)
\end{multline}
where $\Delta_i=\frac{\omega^2-\omega_i^2}{2\omega}$ and $G_i=\lambda/4\omega$.
The KB averaging method has been shown to be the classical limit of the quantum van Vleck Floquet expansion~\cite{seibold2024floquetexpansioncountingpump, VenkatramanStatic2022,burshteinHamiltonian1962}. As such, the same effective slow-flow EOMs can be derived from the semi-classical limit of the quantum Hamiltonian in Eq.~\eqref{eq: RWA Qham}. By doing this, one can show that $\expval{\hat{a}} = \sqrt{\omega /2 \hbar } (u-i v)$, so that the quadratures are the real and imaginary components of the complex coherent state. More details are provided in Appendix Sec.~\ref{sec: relation Q and C}.

We investigate the long-term dynamics (stationary states) of the single mode $i$ by setting $\dot{u}_i=\dot{v}_i=0$ in Eqs.~\eqref{eq: slow-flow equations1}-\eqref{eq: slow-flow equations2}, and solving the resulting coupled polynomial equations~\cite{kosata2022harmonicbalance}. Based on the number of real (physical) solutions, their values, and their stability, we can define a phase diagram for the system, see Fig.~\ref{fig: Fig_1}(c). The region where the zero-amplitude solution is unstable corresponds to the parametric instability condition~\eqref{eq: instability_condition_lyapunov}. Computing the real part of the eigenvalues of the Jacobian of the slow-flow equations [Eqs.~\eqref{eq: slow-flow equations1}-\eqref{eq: slow-flow equations2}] for the zero-amplitude solutions yields
\begin{equation} \label{eq: lyapunov eigenvalues}
	\mu^{(i)}_{\pm}=\frac{-2 \gamma  \omega \pm\sqrt{\lambda ^2-4 \left(\omega ^2-\omega_i^2\right)^2}}{4 \omega }\,,
\end{equation}
which correspond to the Lyapunov exponents. Applying the condition~\eqref{eq: instability_condition_lyapunov} yields the parametric instability lobe (up to order $\epsilon$):
\begin{equation} \label{eq: stability condition}
	\lambda^{(i)}(\omega) = 2\sqrt{\gamma_i^2\omega^2 + (\omega^2-\omega^2_i)^2}\,.
\end{equation}

Below the parametric threshold $\lambda^{(i)}_\mathrm{th}=2\gamma_i\omega_{i}$, the zero-amplitude solution remains the sole stable stationary solution.
Similarly, when $\alpha_i>0$ ($\alpha_i<0$) and the drive is far blue-(red-)detuned, $\omega<\omega_i$ ($\omega>\omega_i$), the zero-amplitude solution is the only stable stationary solution; we mark this region as I in Fig.~\ref{fig: Fig_1}(c).
Within the instability lobe [region II in Fig.~\ref{fig: Fig_1}(c)], the zero-amplitude solution becomes unstable [cf.~gray line in Fig.~\ref{fig: Fig_1}(c)]. Due to the parametric drive, the system then experiences a spontaneous $\mathbb{Z}_2$ time-translation symmetry-breaking phase transition~\cite{ParametricLeuch2016,SorienteDistinctive2021}, or period-doubling bifurcation, resulting in two stable states [cf.~blue and maroon lines in Fig.~\ref{fig: Fig_1}(c)] with amplitudes~\footnote{The magnitude of $\alpha_i$ only changes the amplitude of the phase states, as it can be eliminated by rescaling $x_i$ with $\sqrt{\alpha_i}$.}
\begin{equation} \label{eq: amplitude slow_flow}
	A_{i,\sigma}= \sqrt{\frac{2(\sqrt{\lambda^2-4\gamma^2\omega^2} + 2\omega^2 - 2\omega_{i}^2)}{3\alpha_i}}\,,
\end{equation}
with $\sigma\in \{-1,1\}$ iterating over the two period-doubled states. These states --- also called phase states --- are degenerate in amplitude, but are $\pi$-shifted in  phase, i.e., $\phi_{i,\sigma}=\phi_{i,-\sigma}+\pi$. The phase states persist after leaving the parametric instability lobe for $\alpha_i>0$ ($\alpha_i<0$) and $\omega>\omega_i$ ($\omega<\omega_i$), while the zero amplitude solution regains stability [region III in Fig.~\ref{fig: Fig_1}(c)]. Note that as we are dealing with out-of-equilibrium phase transitions alongside so-called phase states, we will use in the following the mathematical term ``bifurcation'' to denote the boundaries at which phase transitions occur. This choice aims to prevent any ambiguity associated with the multiple meanings of `phase' in this context.

%%%%%%%%%%%%%%%%%%%%%%%%%%%%%%%%%%%%%%%%%%%%%%%%%%%%%%%%%%%%%%%%%%%%%%
\begin{figure}[tb]
	\centering
	\includegraphics[width=1.0\linewidth]{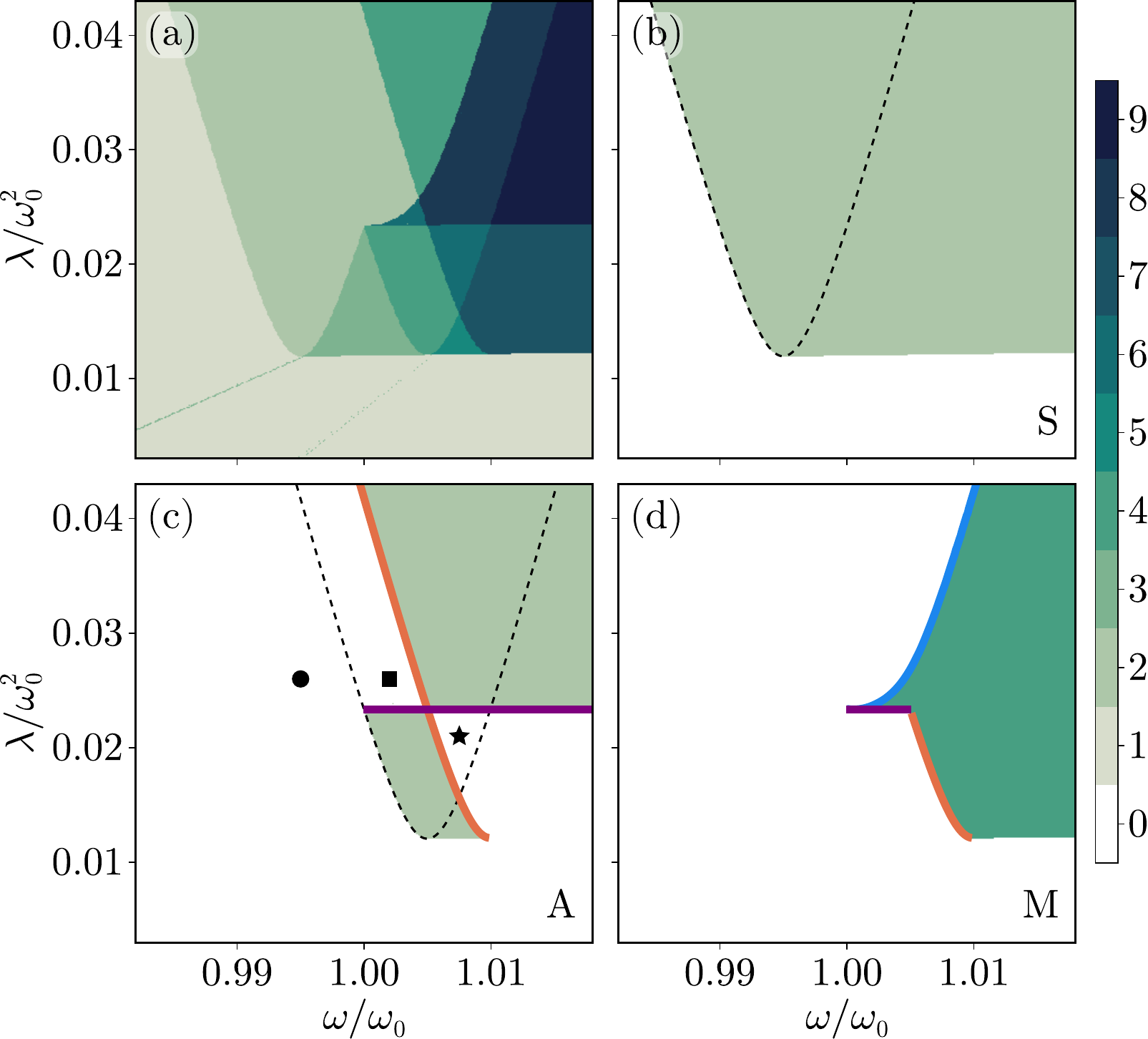}
	\caption{Phase diagram of two homogeneously coupled KPOs [cf. Eqs.~\eqref{eq: eom_normal_mode_2_sym} and~\eqref{eq: eom_normal_mode_2_antisym} and Appendix~\ref{sec: 2 id KPO}], as solved by HarmonicBalance.jl~\cite{kosata2022harmonicbalance}. (a) Number of stable states as a function of the dimensionless parametric pump amplitude $\lambda/\omega_0^2$ and driving frequency $\omega/\omega_0$ with $\gamma/\omega_0=0.006$ and $J/\omega_0^2=0.01$. The states can be classified into (b)~symmetric, (c)~antisymmetric, and (d)~mixed-symmetry states, yielding their own respective stability phase diagrams. Dashed lines mark the Arnold tongue of the symmetric and antisymmetric mode in the absence of cross-Kerr terms in (b) and (c), respectively. Solid colored lines mark the modified parametric instability lines due to cross-Kerr coupling, cf.~Eqs.~\eqref{eq: mod_eom_normal_mode_2_sym} and~\eqref{eq: bif lines symmetric state}. Black square and circle mark the parameters used in Fig.~\ref{fig: Fig_3_extra}.\label{fig: Fig_2}}
\end{figure}

\subsection{The decoupled network state space} \label{sec: state space}
In the decoupled network ($J_{ij}=0$), we have seen that each mode can have up to three different stable solutions. We can formally map the two phase states of the KPO to two spin states $\n$ and $\s$~\cite{goto2016bifurcation,Puri_2017,DykmanInteraction2018,HeugelIsing2022}. Specifically, we denote
\begin{equation} \label{eq: spin bare basis notation}
	\ket{\sigma_i}=A_{i,\sigma}e^{i \phi_{i,\sigma} },
\end{equation}
where the $\sigma_i$ will map onto to $\n, \s$. Note that we use here coherent states' Dirac notation $\ket{...}$ to mark the semiclassical stationary amplitudes, cf.~Eq.~\eqref{eq: amplitude slow_flow}. Analogously, with the combination of the zero-amplitude state, $\ket{0}$, we can label the stable stationary solution space of the KPO as a spin-1 state space.
As indicated in Fig.~\ref{fig: Fig_1}(a), we will represent this set of states as $s=\{\n,\s, 0\}$, where $\n$ and $\s$ represent the phase states, while $0$ signifies the zero-amplitude solution [cf. Fig.~\ref{fig: Fig_1}(c)]. As we have $N$ modes in the decoupled network, we can have up to $3^N$ stable stationary solutions. This is in conjunction with saturating the upper limit of possible stable stationary solutions in such systems~\cite{stableComm,BreidingAlgebraic2022,borovik2023khovanskii,danzlWeakly2010}. As such, we can label the states of the decoupled network with a tensor basis $\mathcal{J}$, represented by $\ket{\psi}_{\mathcal{J}}=\ket{s_1\ldots \, s_N}_{\mathcal{J}}$, where we keep in mind that not all states are accessible at the different I-III regions of parameter space, cf.~Fig.~\ref{fig: Fig_1}(c).

\subsection{Coupled KPO network} \label{sec: nonlinear terms}
When treating the full nonlinear coupled KPO network ($J_{ij}\neq0$), the normal mode basis  remains useful, cf.~Eq.~\eqref{eq: normalEOMs}, alongside the introduction of some challenges. Specifically, the nonlinear terms in the bare basis transform to
\begin{multline}\label{eq: nonlinear terms}
	\alpha C_k x_k^3 +\alpha \sum_{l\neq k}(C_{lk} x_l^2x_k)\\+\alpha \sum_{m\neq l\neq k}(C_{lm} x_l^2 x_m+C_{lmk}x_lx_kx_m)\,,
\end{multline}
for the $k^{\rm th}$ mode in the normal mode basis, where the $C$-coefficients are determined by the transformation $\vb{U}$. We observe that the contributions can be split into self-nonlinear terms $x_k^3$ and cross-nonlinear terms, where the latter are also known as wave-mixing terms. Crucially, the self-nonlinear term stabilizes the individual normal modes as they undergo a parametric phase transition, i.e., they bifurcate into their phase states at their respective parametric instability threshold, cf.~Eqs.~\eqref{eq: normalEOMs}-\eqref{eq: stability condition} with $i \rightarrow k$. This allows us to define a tensor basis ${\mathcal{N}}$ spanned by $\ket{\Psi}_{\mathcal{N}}=\ket{S_1\ldots \, S_N}_{\mathcal{N}}$, where the effective spin $S=\{\n,\s, 0\}$ is based on the stable stationary solutions of the normal modes. Hence, without nonlinear coupling (wave mixing) terms between the normal modes, the phase diagram of the network is composed of parametric instability lobes centered around $\nu_k$, in similitude to the $J_{ij}=0$ case, cf.~Figs.~\ref{fig: Fig_1}(b) and (c). The tensor basis $\mathcal{N}$ still saturates the maximum allowed $3^N$ stable stationary solutions~\cite{BreidingAlgebraic2022,borovik2023khovanskii,stableComm}. Crucially, thus the state of the fully coupled network can be spanned either by the $\ket{\psi}_{\mathcal{J}}$ or the $\ket{\Psi}_{\mathcal{N}}$ basis.

By direct solution of the coupled network equations as well as in experiments, we observe that in addition to the expected bifurcations into phase states for the normal modes, the cross-nonlinear terms between normal modes [cf. Eq.~\eqref{eq: nonlinear terms}] can induce additional structure to the phase diagram, see Fig.~\ref{fig: Fig_2}(a) and Refs.~\cite{HeugelIsing2022,heugel2023proliferation,heugel2023role}. Once the amplitudes of the normal modes $\alpha A_k^2$ become large [cf.~Eq.~\eqref{eq: amplitude slow_flow} with $i \rightarrow k$], the cross-nonlinear coupling between the normal modes can lead to new states of mixed symmetry, having an amplitude contribution in multiple normal modes~\cite{HeugelIsing2022}. Furthermore, the nonlinear coupling influences the stability of the normal-mode phase states, leading to clear deviations from the naive parametric lobe picture~\cite{HeugelIsing2022,heugel2023role} with profound implications in the realm of stochastic dynamics~\cite{heugel2023proliferation}.

We have reviewed the phenomenology of coupled KPO networks, while introducing a comprehensive notation for the problem. In the following, we employ a secular perturbation theory on top of our tensor basis, and identify parametric instabilities induced by cross-Kerr coupling. Thus, we develop a physical understanding of how the normal modes interact, which allows us to describe the phase diagram of the KPO network.

\section{Understanding coupled KPO networks: the N=2 case} \label{sec: 2 id KPO}

In accordance with the basis notation developed above, we introduce here a perturbative framework to describe the phase diagram of the coupled KPO network via the stability of the tensor states. Our scheme primarily focuses on the parametric instability thresholds and how they are modified through nonlinear coupling corrections. To illustrate this approach, we proceed to describe the complete phase diagram of two homogeneous KPOs in the following section. Crucially, this simpler example is already providing insights beyond the state of the art~\cite{HeugelIsing2022}. In the subsequent section, we discuss how our approach generalizes to larger networks.

We consider $N=2$ identical and coupled KPOs described by Eq.~\eqref{eq: eom KPO} with  bare frequencies $\omega_i=\omega_0$, parametric drive amplitudes $\lambda_i=\lambda$, damping coefficients $\gamma_i=\gamma$, nonlinearity parameters $\alpha_i=\alpha$, and coupling coefficient $J_{ij}=J$. We henceforth choose both the nonlinearity and the coupling to be positive, whereas analogous analysis will apply for other choices~\footnote{Along the detuning axis, the sign of $J$ interchanges the order in which the normal modes appear, and the sign of $\alpha$ reverses the direction at which the coexistence region appears.}. The EOMs of the system in the normal mode basis read
\begingroup \setmuskip{\medmuskip}{0mu} % thinmuskip or thickmuskip
\begin{align} \label{eq: eom_normal_mode_2_sym}
  \ddot{x}_s +\left[\omega_s^2 - \lambda \cos(2\omega t)\right]x_s + \gamma \dot{x}_s +\alpha(x_s^3+3x_a^2x_s) & =0\,,                                       \\
  \ddot{x}_a +\left[\omega_a^2 -\lambda \cos(2\omega t)\right]x_a + \gamma \dot{x}_a + \alpha(x_a^3+3x_s^2x_a) & =0\,, \label{eq: eom_normal_mode_2_antisym}
\end{align}
\endgroup
where the normal modes comprise a symmetric (S), $x_s=(x_1+x_2)/2$, and an antisymmetric (A) mode, $x_a=(x_2-x_1)/2$, where $\omega_{s/a} = \sqrt{\omega_0^2 \pm J}$. We calculate the full phase diagram from the slow-flow equations derived from Eqs.~\eqref{eq: eom_normal_mode_2_sym} and~\eqref{eq: eom_normal_mode_2_antisym} by setting $\dot{u}_k=\dot{v}_k=0$ and solving for the roots of the coupled polynomials~\cite{kosata2022harmonicbalance}, see Fig.~\ref{fig: Fig_2}(a) and Appendix~\ref{sec: 2 id KPO}. We obtain up to the maximum of $3^2$ possible stable stationary states, but the bifurcation lines differ significantly from the naively expected overlapping parametric lobes of the normal modes, cf.~Fig.~\ref{fig: Fig_1}(b).
\begin{table}[t]
	\renewcommand{\arraystretch}{1.3}
	\centering
	\begin{tabular}{c|c|c}
		\hline
		$N=2$ & \textbf{bare basis} $\mathcal{J}$ & \textbf{normal basis} $\mathcal{N}$
		\\
		\hline
		\textbf{Symmetric}
		      &
		$\state{\n}{\n},\state{\s}{\s}$
		      &
		$\state{\n}{0},\state{\s}{0}$
		\\
		\hline
		\textbf{Antisymmetric}
		      &
		$\state{\n}{\s},\state{\s}{\n}$

		      &
		$\state{0}{\s},\state{0}{\n}$
		\\
		\hline
		\textbf{Mixed}
		      &
		$\state{\n}{0},\state{\s}{0}$, $\state{0}{\s},\state{0}{\n}$
		      &
		$\state{\s}{\n}$, $\state{\n}{\s}$, $\state{\n}{\n}$, $\state{\s}{\s}$
		\\
		\hline
		\textbf{Zero}
		      &
		$\state{0}{0}$
		      &
		$\state{0}{0}$
		\\
		\hline
	\end{tabular}
	\caption{Summary of how the $3^2$ stationary states of a 2-mode KPO network [cf.~Fig.~\ref{fig: Fig_2}] are spanned by the tensor spin-state basis notation, i.e., the bare and normal tensor bases, cf.~Sec.~\ref{sec: state space}. \label{tab: state_classification_2}}
\end{table}

To analyze the discrepancy between Figs.~\ref{fig: Fig_1}(b) and~\ref{fig: Fig_2}(a), we consider the case with only the eigen-Kerr term of each normal mode, i.e., we remove the cross-Kerr terms in Eqs.~\eqref{eq: eom_normal_mode_2_sym} and~\eqref{eq: eom_normal_mode_2_antisym}. In this limit,  the KPO network solutions are spanned by the normal-modes tensor basis $\state{S_s}{S_a}_\mathcal{N}$ with $S_{s,a}$ indicating the stationary state of $x_{s,a}$, respectively. This basis spans over 9 solutions, where each normal mode can reside in each of its three ``spin-1'' states without impacting the state of the other. For example, an S- or an A-phase state in the normal modes basis can read $\state{\n_s}{0}_{\mathcal{N}}$ or $\state{0}{\n_a}_{\mathcal{N}}$, respectively. This limit yields a phase diagram akin to Fig.~\ref{fig: Fig_1}(b) for $N=2$.

When the cross-Kerr terms are present, we obtain Fig.~\ref{fig: Fig_2}(a). We see up to 9 solutions, but they do not directly match to the independent normal mode states described above. To better understand the stationary states that appear in the system, we filter and plot them according to their projection onto the normal mode tensor basis $\mathcal{N}$, see Figs.~\ref{fig: Fig_2}(b) and (c). The S (A) phase states account for four of the stationary states appearing in the system, see Figs.~\ref{fig: Fig_2}(a)-(c). The fifth empty state ($\state{0}{0}_{\mathcal{N}}$) clearly also manifests.
In addition, however, we observe stationary states with ``mixed'' symmetry that involve combinations of all four states $\ket{S_s\neq 0\,S_a\neq 0}_\mathcal{N}$ [cf.~Eq.~\eqref{eq: spin bare basis notation}], see Fig.~\ref{fig: Fig_2}(d). This observation leads us to suspect that the cross-Kerr terms in Eq.~\eqref{eq: nonlinear terms} play a crucial role in the creation of these states with mixed symmetry.

We recall that the stationary states can alternatively be constructed in the bare mode tensor basis $\state{s_1}{s_2}_{\mathcal{J}}$, where $s_{1,2}\in\{\n,\s,0\}$ denotes a stationary state of oscillator $1$ and $2$, respectively. We can therefore filter the stationary states of the system using the bare tensor basis to find that: the symmetric phase states map onto $\state{\n}{\n}_{\mathcal{J}}$ and $\state{\s}{\s}_{\mathcal{J}}$, while the antisymmetric phase states correspond to $\state{\n}{\s}_{\mathcal{J}}$ and $\state{\s}{\n}_{\mathcal{J}}$. The empty state reads $\state{0}{0}_{\mathcal{J}}$, whereas mixed-symmetry states involve the combination of a phase state from one of the oscillators alongside the zero-amplitude solution of the other, e.g., the states $\state{\n}{0}_{\mathcal{J}}$ and $\state{0}{\n}_{\mathcal{J}}$. Crucially, in this basis, $J$ is the perturbation that couples the nonlinear responses of the bare resonators in order to spawn mixed-symmetry states. We can thus approach the solution space of coupled KPOs either from the normal-mode basis or the bare basis, using different perturbation schemes. The classification of the states in both spin-1 bases is summarized in Table~\ref{tab: state_classification_2}.

In the spirit of these two approaches, we will now analyze the stability of the many-body states within two distinct limits: (i) a limit that is well characterized by normal modes that can perturbatively couple via nonlinear cross-Kerr terms, and (ii) a limit where the parametric drive pins the resonators to a state in the bare mode tensor basis, and the coupling acts as a perturbation. These limits correspond to (i) $\lambda\ll J$ and (ii) $\lambda\gg J$. Together, the two approaches provide a full picture of the system solutions both close to parametric threshold and beyond it.

\begin{figure}[t]
	\centering
	\includegraphics[width=1.0\linewidth]{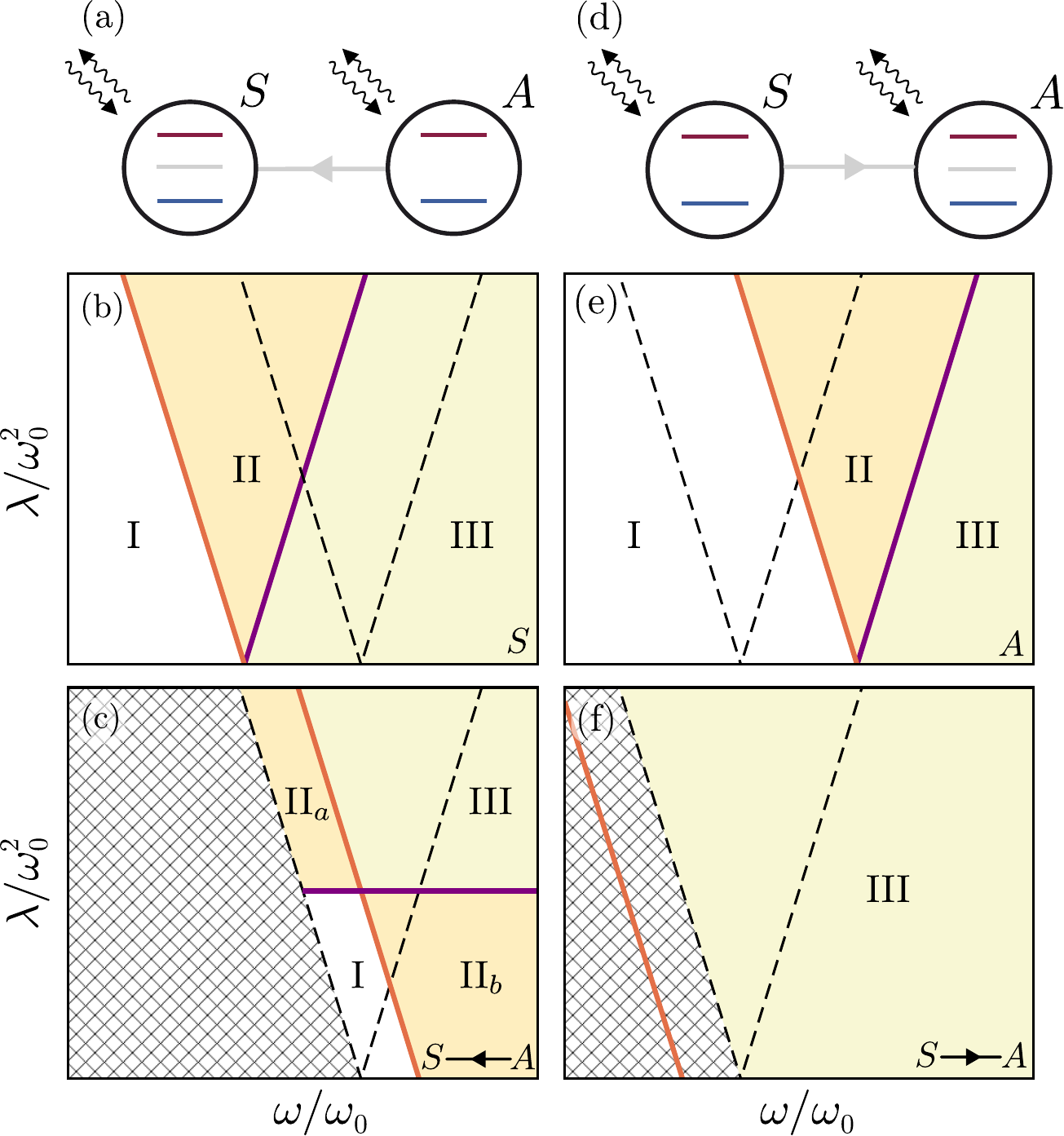}
	\caption{Influence of the cross-nonlinear coupling between the symmetric (S) and antisymmetric (A) normal modes of two homogeneously coupled KPOs with $J>0$ and $\alpha>0$ [cf.~Eqs.~\eqref{eq: eom_normal_mode_2_sym} and~\eqref{eq: eom_normal_mode_2_antisym}]. (a)~and (d)~Illustration of the directional perturbative schemes of how the A (S) mode [circles with lines marking spin-1 states] parametrically drives [directional arrow] the S (A) mode [cf.~Eqs.~\eqref{eq: mod_eom_normal_mode_2_sym}]. (b)~and (e)~Phase diagram of the uncoupled S (A) mode. The white (I), orange (II), and yellow (III) regions denote the region, where only the zero-amplitude state, only the phase states are stable, or both are stable, for the given normal mode, respectively. The black dashed lines indicate the parametric lobe (bifurcation lines) of the other mode. (c)~and (f)~State-dependent stability diagram for the S (A) mode perturbed by an A (S) oscillation, cf.~Eq.~\eqref{eq: mod_eom_normal_mode_2_sym}. Inside the hashed region, the A (S) mode can only have a zero amplitude, hence, no state-dependent stationary states of S (A) exist. The orange and purple lines indicate the state-dependent bifurcation lines of the S (A) mode, cf.~Eq.~\eqref{eq: bif lines symmetric state} [\eqref{eq: bif lines antisymmetric state}].
	}
	\label{fig: Fig_3}
\end{figure}

\subsection{$\lambda\ll J$} \label{sec: lambda<J}
In the limit of $\lambda\ll J$, we treat the modes as strongly coupled, such that they are hybridized into normal modes.
Neglecting the nonlinear coupling terms $x_i^2x_j$, we obtain two normal modes acting as decoupled KPOs with an expected phase diagram that superimposes the respective I-III regions of each of the two non-interacting KPOs, cf. Figs.~\ref{fig: Fig_1}(b) and (c), as well as dashed lines in Fig.~\ref{fig: Fig_2}.
However, when including the cross-nonlinear coupling terms, we observe a host of additional features, cf.~Fig.~\ref{fig: Fig_2}, which we will resolve analytically in the following.

We perturbatively include the impact of the nonlinear coupling between the normal modes.
%For instructive purposes, we use a naive approach to include the impact of nonlinear coupling between the normal modes. 
Our method adapts ideas from secular perturbation theory while including the stationary nonlinear motion of a mode, following the KB averaging method. For the sake of simplicity, we neglect damping in the following analysis. A more rigorous treatment using secular perturbation theory that incorporates dissipation is presented in Appendix~\ref{sec: secular perturbation}.

\subsubsection{Perturbation of the S-mode by the A-mode}
Let us start by analyzing the impact of the stationary A-mode's motion on the S-mode, see Fig.~\ref{fig: Fig_3}(a).
%\footnote{Changing the sign of the coupling J revert the two normal modes, making the antisymmetric the high amplitude mode and symmetric the low amplitude mode.}.
From KB averaging, we can assume that the A-mode oscillates at only one frequency $\omega$ as $x_a = A_a \cos(\omega t +\phi_a)$, with amplitude $A_a$ and phase $\phi_a$. When dissipation is neglected, the phase of the stationary mode locks to the parametric drive such that $\phi_a=0$ or $\phi_a=\pi$ [cf. Eq.~\eqref{eq: eom KPO}].
Substituting the expression into the EOM of the S-mode~\eqref{eq: eom_normal_mode_2_sym}, we obtain a modified EOM
\begin{equation} \label{eq: mod_eom_normal_mode_2_sym}
	\ddot{x}_s+\left[\omega_s^2 +\frac{3\alpha}{2}A_a^2+\left(\frac{3\alpha}{2}A_a^2-\lambda\right)\cos(2\omega t)\right]x_s + \alpha x_s^3=0\,,
\end{equation}
of the same form as a single KPO, where the A-mode induces detuning to the S-mode's bare frequency (an effect akin to the AC Stark shift~\cite{cohen1986quantum}), as well as \textit{parametrically} drives the S-mode~\cite{denardoParametric1999}.
% Note that in the presence of dissipation, the response phase $\phi_a$ of the A-mode rotates away from the origin ($+\pi$), see Appendix~\ref{dissipation}.

The stationary solutions of Eq.~\eqref{eq: mod_eom_normal_mode_2_sym} for the {S-mode} depend on which stationary motion the {A-mode} has, i.e., what value $A_a$ takes. Consequently, we dub them \textit{state-dependent stationary states}, cf.~Appendix~\ref{sec: secular perturbation}. We recall that, to first-order KB, the A-mode (in the absence of backaction from the S-mode) has three (I-III) different possible regions depending on the mode's stability, cf.~Eq.~\eqref{eq: stability condition} with $\omega_i\to\omega_a$ and $\alpha_i\to\alpha/4$, as well as Fig.~\ref{fig: Fig_1}(c). Depending on the region and nonlinear oscillation solution, $A_a$ can either be zero or lock into one of its phase states. In the former case, Eq.~\eqref{eq: mod_eom_normal_mode_2_sym} returns to describe the bare {S-mode} dynamics without the influence of the A-mode, see Fig.~\ref{fig: Fig_3}(b). In the latter, we plug in the non-trivial amplitude solution given by Eq.~\eqref{eq: amplitude slow_flow} into Eq.~\eqref{eq: mod_eom_normal_mode_2_sym}. We thus obtain a modified Mathieu equation that has a shifted instability lobe [cf. Eqs.~\eqref{eq: instability_condition_lyapunov} and~\eqref{eq: stability condition}]. We calculate the state-dependent bifurcation lines for the perturbed symmetric state to be
\begin{equation}\label{eq: bif lines symmetric state}
	\lambda^{(s)}_\mathrm{1}
	=2J
	\qq{and}
	\lambda^{(s)}_\mathrm{2}=4 J-2 \omega ^2+2 \omega_0^2\,,
\end{equation}
and draw them in Fig.~\ref{fig: Fig_3}(c) as purple and orange lines, respectively.
We obtain these lines using the same approach as that used for Eq.~\eqref{eq: stability condition}, i.e., by employing the KB averaging to the lowest order of Eq.~\eqref{eq: mod_eom_normal_mode_2_sym} and analyzing when the Jacobian of the slow-flow equation is equal to zero [cf. Eq.~\eqref{eq: lyapunov eigenvalues} with $\omega_i\to \omega_s^2+\frac{3\alpha}{2}A_a^2$ and $\lambda_i\to \frac{3\alpha}{2}A_a^2-\lambda$].

It is important to note that the validity of the latter perturbed phase diagram relies on the A-mode residing in a nonzero amplitude state. We dub such a perturbed stability diagram a \textit{state-dependent stability diagram}, because it is the outcome of a conditional perturbation that depends on the state of the perturbing system. In other words, the state-dependent stability diagram is a copy of the unperturbed diagram, but shifted depending on $A_a$. We mark in Fig.~\ref{fig: Fig_3}(c) the region where the state-dependent perturbation cannot manifest. To summarize, the phase transition lines captured by Eqs.~\eqref{eq: bif lines symmetric state} can be attributed to the A-mode impacting the S-mode. Comparing with Fig.~\ref{fig: Fig_2}, these transition lines manifest in the full phase diagram of the system, as indicated by matching colored boundaries. In the presence of damping, these lines are curved similarly as in Eq.~\eqref{eq: stability condition}, and as shown numerically in Appendix~\ref{sec: extr info 2}.

The state-dependent instability lines distinguish between regions that are stable under the conditional parametric drive, and those that become parametrically unstable, see in Fig.~\ref{fig: Fig_3}(c). Similar to the standard KPO case [cf.~Fig.~\ref{fig: Fig_1}(c)], stability to the conditional parametric drives leads to a region I$_{S\leftarrow A}$ (white) where the S-mode can only have the zero solution, and region III$_{S\leftarrow A}$ (yellow) where the zero S-mode can coexist with its phase states. Instability is ``within'' the lobe (orange), which could mean two things: either the ansatz that the A-mode is in a phase state breaks down, or the A-mode drives the S-mode into parametric phase state, which would result in combined stationary states with mixed-symmetry, e.g., $\state{\s}{\n}_{\mathcal{N}}$. In Sec.~\ref{sec: lambda>J}, we will be able to distinguish between the two cases: a region II$_{a,S\leftarrow A}$ where the S-mode becomes parametrically unstable and eats up the gain from the A-mode, thus, causing the A-mode to lose its high-amplitude state; and region II$_{b,S\leftarrow A}$ where both the S- and A-mode can populate their phase states in parallel; these will be mixed-symmetry states, as discussed in Sec.~\ref{sec: lambda>J}.

Our state-dependent procedure implies that the A-states $\state{0}{\n}_{\mathcal{N}}$ and $\state{0}{\s}_{\mathcal{N}}$ are unstable in regions II in Fig.~\ref{fig: Fig_3}(c) [cf.~Fig.~\ref{fig: Fig_2}(c)], contrary to the naive expectation. We can visualize how they become unstable using a numerical ringdown experiment, see Fig.~\ref{fig: Fig_3_extra}. In Fig.~\ref{fig: Fig_3_extra}(a), we set the system in an antisymmetric configuration within region II of the S-mode but where the A-mode can only be zero, cf.~Fig.~\ref{fig: Fig_3}(b) and (c). As the system relaxes to its chosen stationary state, the A-mode immediately rings down to zero amplitude. At the same time, the S-mode stabilizes to a finite amplitude independent of the A-mode, indicating that the symmetric mode is only driven by the global parametric drive. Instead, if we initialize the system in the same antisymmetric configuration within region II$_{a,S\leftarrow A}$, different dynamics unfold, see Fig.~\ref{fig: Fig_3_extra}(b). Here, the antisymmetric mode initially decays into a finite amplitude (a metastable state). This finite amplitude of the A-mode starts to parametrically amplify the S-mode. As the S-mode gains amplitude, the A-mode loses its stability, yielding a symmetric phase state. We will discuss the dynamics in Fig.~\ref{fig: Fig_3_extra}(c) after we resolved the physics of region II$_{b,S\leftarrow A}$ in Sec.~\ref{sec: lambda>J}.

\subsubsection{Perturbation of the A-mode by the S-mode}
We can also conduct the state-dependent perturbation for how the S-mode influences an A-mode [cf.~Fig.~\ref{fig: Fig_3}(d)]. We assume that the S-mode locks onto one of its phase states, in regions II or III of Fig.~\ref{fig: Fig_3}(e). Then, we derive a shifted Mathieu equation analogous to Eq.~\eqref{eq: mod_eom_normal_mode_2_sym} (with $s$ and $a$ interchanged). This analysis results in the phase diagram shown in Fig.~\ref{fig: Fig_3}(f), characterized by the bifurcation lines:
\begin{equation}\label{eq: bif lines antisymmetric state}
	\lambda^{(a)}_\mathrm{1} = -2J,
	\qq{and}
	\lambda^{(a)}_\mathrm{2} = -4 J - 2 \omega^2 + 2 \omega_0^2.
\end{equation}
Similar to Eq.~\eqref{eq: bif lines symmetric state}, these bifurcation lines, $\lambda^{(a)}_\mathrm{1}$ and $\lambda^{(a)}_\mathrm{2}$, delineate the conditions under which the A-mode becomes parametrically excited depending on the S-mode being in a parametric phase state. Unlike in Fig.~\ref{fig: Fig_3}(c), the lines in Fig.~\ref{fig: Fig_3}(f), do not impose new partitions of the stability diagram. We are left with only a state-dependent region III, where is appears that ``everything goes'', i.e., by our current treatment, all possible combinations of the normal-mode tensor basis can manifest.

\subsection{$\lambda\gg J$} \label{sec: lambda>J}
\begin{figure}[t]
	\centering
	\includegraphics[width=\linewidth]{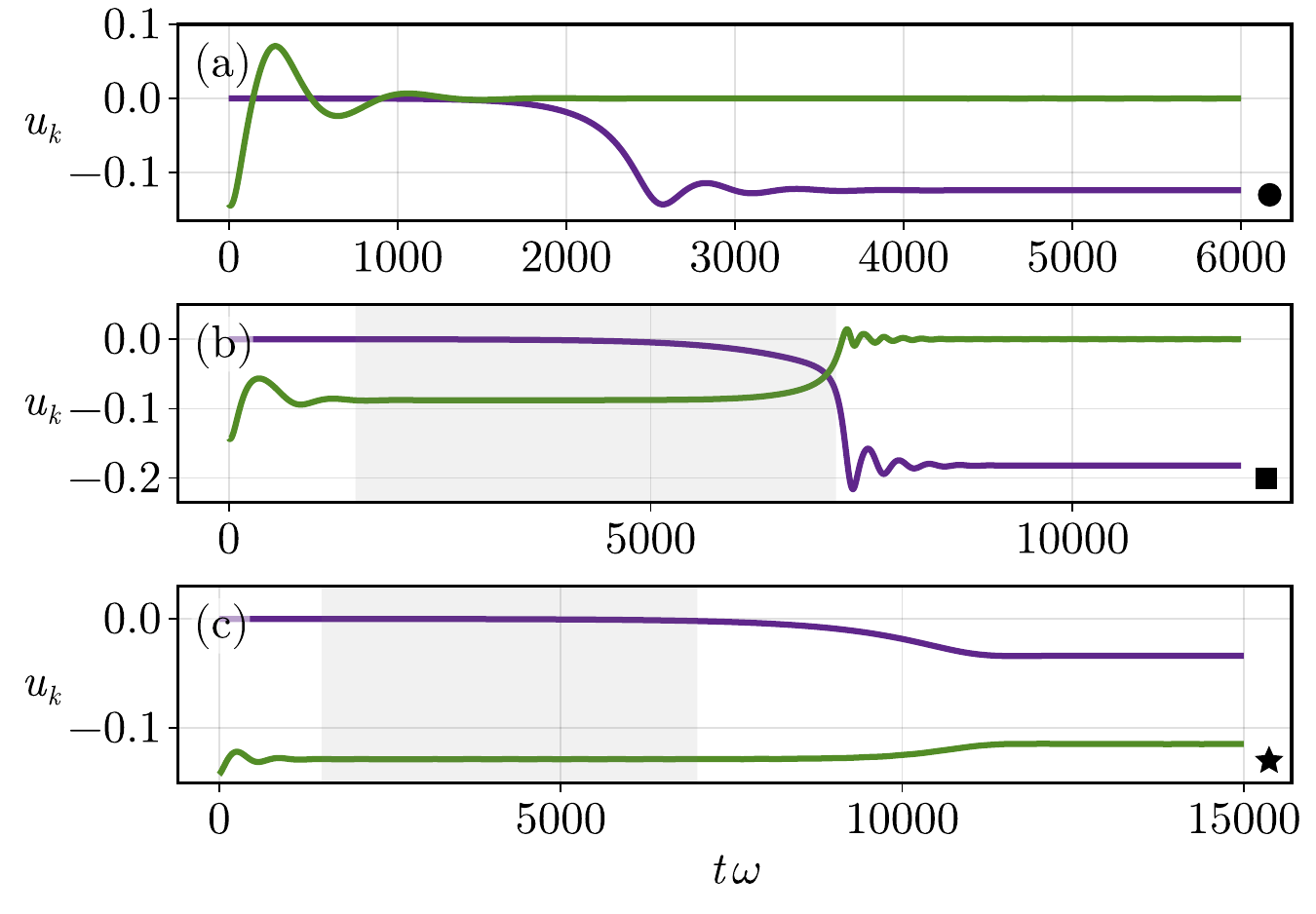}
	\caption{Temporal evolution of the quadrature $u$ of the symmetric (S) and the antisymmetric (A) modes, starting from the same antisymmetric initial condition. (a) Outside the parametric interaction region [cf.~circle marker in Fig.~\ref{fig: Fig_2}], the antisymmetric state $\state{0}{\n}_{\mathcal{N}}$ quickly decays while the symmetric mode rings up to a finite amplitude. (b) Inside the parametric interaction region II${}_a$ [cf.~square marker in Fig.~\ref{fig: Fig_2}], the antisymmetric state persists (gray region) before decaying due to the parametric amplification of the symmetric mode. (c) Inside the parametric interaction region II${}_b$ [cf.~star marker in Fig.~\ref{fig: Fig_2}], the antisymmetric state persists (gray region) while ringing up the symmetric mode, yielding a mixed-symmetry stationary state. }
	\label{fig: Fig_3_extra}
\end{figure}

In our perturbative analysis, depicted in Fig.~\ref{fig: Fig_3}, we captured two bifurcation lines (orange and purple) within the  phase diagram of the fully coupled system illustrated in Fig.~\ref{fig: Fig_2}(a). Nevertheless, the mixed states, as shown in Fig.~\ref{fig: Fig_2}(d), appear to be influenced by an additional bifurcation line, marked in blue, which govern their stability transitions. To elucidate the nature of this bifurcation line, we examine the regime where $\lambda\gg J$, i.e., the scenario in which the parametric drive dominantly determines the resonators' states within the bare mode tensor basis, and the coupling is merely a small perturbation.

In the case of negligible coupling $J$ in Eq.~\eqref{eq: eom KPO}, we have two degenerate parametric bare modes with coinciding parametric phase diagrams. The bifurcation lines between the different I-III regions are indicated by Eq.~\eqref{eq: stability condition}, and are centered precisely around $\omega_0$:
\begin{equation}\label{eq: bif line bare modes}
	\lambda^{(1/2)}_\pm(\omega) = \pm 2|\omega^2-\omega_0^2|\,.
\end{equation}
As discussed in Section \ref{sec: section 2}, the mixed-symmetry states result from a zero-amplitude bare mode coupled to a phase state of another bare mode, e.g., $\state{\n}{0}_{\mathcal{J}}$ or $\state{0}{\s}_{\mathcal{J}}$. Consequently, mixed-symmetry states cannot appear  when both resonators are subject to parametric excitation, i.e., cannot manifest in region II here. Crucially, the bifurcation lines in Eq.~\eqref{eq: bif line bare modes} are only valid in the regime where $\lambda > \lambda^{(s)}_1=\omega_a^2-\omega_s^2 =2J$, as this marks the threshold where the parametric drive is strong enough to overcome the energy separation between the normal modes imposed by the coupling $J$. Below this threshold, the system's dynamics are dominated by the normal mode structure, and the bare mode description becomes inadequate.

The bifurcation line $\lambda^{(1/2)}_+$ separates the regions II$_{a,S\leftarrow A}$ and II$_{b,S\leftarrow A}$ in Fig.~\ref{fig: Fig_2}(c)elow $\lambda^{(1/2)}_+$, a mixed-symmetry state, e.g., $\state{\n}{0}_{\mathcal{J}}$, is stable. In addition, in region II$_{b,S\leftarrow A}$ (or III$_{S\leftarrow A}$), the same state $\state{\n}{0}_{\mathcal{J}}\simeq \state{\s}{\n}_{\mathcal{N}}$ is stable due to parametric interaction between the A- and S-modes. We can visualize how the dynamics manifests in this region in Fig.~\ref{fig: Fig_3_extra}(c), by initializing the system at the star marker in II$_{b,S\leftarrow A}$. Here, the A-mode parametrically excites the S-mode but retains its finite amplitude, yielding a mixed-symmetry state.

\begin{figure*}[t]
	\centering
	\includegraphics[width=1.0\linewidth]{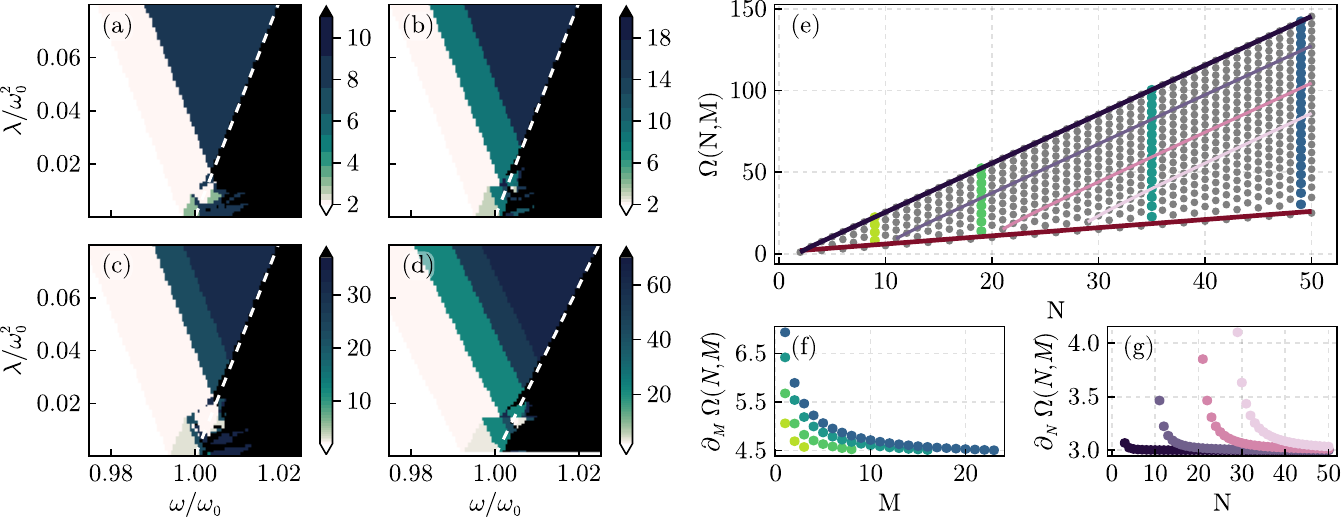}
	\caption{
		Stability phase diagrams and frequency shifts for homogeneously coupled KPO networks. (a)-(d) Stability phase diagrams for networks of 3 to 6 coupled KPOs, respectively as solved by HarmonicBalance.jl~\cite{kosata2022harmonicbalance}. It shows the number of stable states as a function of the dimensionless parametric pump amplitude $\lambda/\omega_0^2$ and driving frequency $\omega/\omega_0$ with $\gamma/\omega_0=0$ and $J/\omega_0^2=0.003$. The color scale is capped at $2^N$ to highlight the regions where the number of stable states matches the number of Ising states in a network with $N$ nodes. The white dashed lines represent the bifurcation line $\lambda^{(1/2)}(\omega)$ from Eq.~\eqref{eq: bif line bare modes}. (e) Frequency shift $\Omega(N, M)$ (cf.~Eq.~\eqref{eq: shift}, gray dots) for every partition $\{n=\frac{1}{2}(N+M), m=\frac{1}{2}(N-M)\}$ up to $N=50$. The trend lines for partitions $\{N-9, 9\}$, $\{N-19, 19\}$, and $\{N-27, 27\}$ are indicated in increasingly lighter shades of purple. (f) Difference in frequency shift $\Omega(N, M)$ with respect to $M$ for network sizes $N=9, 19, 35, 49$, showing a convergence to $9/2$, cf.~Eq.~\eqref{eq: shift limit 1}. (g) Difference in frequency shift $\Omega(N, M)$ with respect to $N$, indicating a uniform spacing between bifurcation lines in the large $N$ limit, cf.~Eq.~\eqref{eq: shift limit 2}. \label{fig: Fig_5}
	}
\end{figure*}

\section{N identical all-to-all coupled KPO network} \label{sec: N id J}

In this section, we employ the insights from Sec.~\ref{sec: 2 id KPO} to extend our analysis to networks of $N$ identical KPOs with all-to-all coupling. Specifically, we show how the bifurcation lines appearing for the $N=2$ case manifest in larger networks. Thus, we reveal a cascade of parametric instabilities with regions that can be understood by partitioning the stationary states of the network. Through this analysis, we obtain analytical expressions for the bifurcation lines, which enable us to understand the phase diagram of the network. Crucially, we find that in the large-$N$ limit, these bifurcation lines become uniformly spaced, leading to a predictable structure. This analysis has important implications for the implementation of KPO networks as Ising machines.
% A similar analysis can be conducted for other highly symmetric coupling topologies. For example, the case of a ring topology is discussed in the appendix~\cite{supmat}.

As an illustration, we first solve for all the stationary states of the network with $N=3,\ldots,6$ using HarmonicBalance.jl~\cite{kosata2022harmonicbalance}, see Figs.~\ref{fig: Fig_5}(a)-(d). As the network is all-to-all equally coupled, we expect that in the absence of cross-Kerr coupling, parametric lobes would appear for every normal mode; one for the symmetric states of the system and $N-1$ degenerate antisymmetric lobes, cf.~Sec.~\ref{sec:stability linear}. In the phase diagram with cross-Kerr coupling, however, we see a cascade of $\left\lfloor \frac{N+2}{2} \right\rfloor$ shifted instability lines, akin to the orange line in Fig.~\ref{fig: Fig_2}.

To understand this cascade of lines, we focus on the stability of network states where all the individual bare modes are excited into phase states, i.e., states that do not have zero entries in their bare tensor basis decomposition, e.g., $\ket{\n\s\n \cdots \s}_{\mathcal{J}}$. We coin such network states ``Ising states'', because they most closely resemble a network of Ising spins. In order to limit the system to Ising states, all the states involving a zero-amplitude bare mode in the decomposition have to be unstable. In the following, we refer to such states as `non-Ising states'.

Similar to the case with $N=2$ in section~\ref{sec: lambda>J}, in the limit of $\lambda\gg J$, the stability of non-Ising states is delineated by the bifurcation line defined in Eq.~\eqref{eq: bif line bare modes}, cf.~blue line in Fig.~\ref{fig: Fig_2}. We draw this bifurcation line $\lambda^{(1/2)}_+$ on top of the numerical stability phase diagrams in Figs.~\ref{fig: Fig_5}(a)-(d). Above that line, non-Ising states are unstable, and the system is composed only of Ising states. Recall from Sec.~\ref{sec: lambda>J} that the line $\lambda^{(1/2)}_+$ relates to a scenario when the bare modes are decoupled; it marks that states involving the zero-amplitude modes are only stable outside the parametric lobe centered around the bare frequency $\omega_0$.

In the region bounded by the bifurcation line $\lambda^{(1/2)}_+$, we expect the Ising states to become stable when increasing the parametric drive. en each Ising state undergoes a bifurcation in the cascade, we partition the Ising states depending on the absolute difference $M=|n-m|$ between the number of spins pointing up $n$ and down $m$. We denote such a partition as $\{n,m\}$, where $n+m=N$. We only care about the absolute difference $M$ due to the $\mathbb{Z}_2$ symmetry between phase states. As such, the partitions $\{n,m\}=\{m,n\}$, and we will refer to both as $\{n,m\}$. In Table~\ref{tab: partitions}, we summarize the partitions until $N=7$. With increasing number of modes in the network, the number of partitionsgrows by $\left\lfloor \frac{N+2}{2} \right\rfloor$. The partitions can additionally be labeled by their normal mode symmetry, cf.~Sec.~\ref{sec:stability linear}.

Generally, we have $N$ coupled EOMs for the system. Nonetheless, when we consider a partition $\{n,m\}$ and plug it into the EOMs~\eqref{eq: eom KPO}, we obtain two sets of identical EOMs, one set with $n$ identical EOMs and another with $m$. The coupling between the two sets of EOMs leads to two reduced EOMs for the network
\begingroup
\medmuskip=0.1mu
\thinmuskip=0.1mu
\thickmuskip=0.1mu
\begin{align} \label{eq: eom_reduced}
	 & \resizebox{.87\hsize}{!}{$\displaystyle \ddot{x}_n + [\omega_{0}^2-(m-1)J-\lambda\cos(2\omega t)]x_n + \alpha x_n^3  - n J x_{m}=0,$}     \\
	 & \resizebox{.87\hsize}{!}{$\displaystyle \ddot{x}_{m} + [\omega_{0}^2-(n-1)J-\lambda\cos(2\omega t)]x_{m} + \alpha x_{m}^3  - m J x_n=0,$}
\end{align}
\endgroup
where $x_n$ and $x_m$ represent the displacement of the $n$ and $m$ resonators. Given that we now effectively have two variables, it is convenient to use an effective normal mode basis: $x_+=x_n+x_m$ and $x_-=x_n-x_m$. The new EOMs read in this basis
\begingroup
\medmuskip=0.1mu
\thinmuskip=0.1mu
\thickmuskip=0.1mu
\begin{align} \label{eq: normal mode all-to-all1}
	 & \resizebox{.87\hsize}{!}{$\displaystyle\ddot{x}_+ +\left[\omega_+^2 - \lambda \cos(2\omega t)\right]x_+ + \alpha( x_+^3 + 3x_-^2x_+)-MJ x_-=0$}\,, \\
	 & \resizebox{.73\hsize}{!}{$\displaystyle\ddot{x}_- +\left[\omega_-^2 - \lambda \cos(2\omega t)\right]x_- + \alpha( x_-^3 + 3 x_+^2x_-)=0$}\,.
	\label{eq: normal mode all-to-all2}
\end{align}
\endgroup
with $\omega_+^2=\omega_0^2-(N-1)J$, and $\omega_-^2=\omega_0^2+J$. Thus, we find a similar nonlinear coupling structure as for the $N=2$ case [cf. Eqs.~\eqref{eq: eom_normal_mode_2_sym} and~\eqref{eq: eom_normal_mode_2_antisym}], however, with an additional non-reciprocal coupling where the (antisymmetric) $x_-$ mode impacts the (symmetric) $x_+$ mode.

Applying KB perturbation to Eqs.~\eqref{eq: normal mode all-to-all1} and~\eqref{eq: normal mode all-to-all2}, we can compute the stability of the partition $\{n,m\}$ which is determined by a standard parametric instability line of the form
\begin{equation}\label{eq: bif shift}
	\lambda_{\{n,m\}}(\omega)=-2(\omega^2-\omega_0^2-J-\Omega(N,M)J)
\end{equation}
with $\Omega(N,M)$ an additional frequency shift. Crucially, every partition has a bifurcation line with the same slope as a function of $\omega$. Different partitions are displaced with respect to one another by the frequency shift
% \begin{widetext}
\begingroup\small
\medmuskip=0.1mu
\thinmuskip=0.1mu
\thickmuskip=0.1mu
\begin{equation}\label{eq: shift}
	\Omega(N,M)= \frac{1}{2}\sqrt{3 \Xi+ \Sigma}+\sqrt{8  \Sigma-12 \Xi-\frac{8  N \left(N^2-9 M^2\right)}{\sqrt{3 \Xi+ \Sigma}}} ,
\end{equation}
\endgroup
% \end{widetext}
with $\Xi=\sqrt[3]{- \left(M^3-M N^2\right)^2}$ and $\Sigma=\left(3 M^2+N^2\right)$. This dependence explains why we have $\left\lfloor \frac{N+2}{2} \right\rfloor$ parallel bifurcation lines above $\lambda^{(1/2)}_+$ in Figs.~\ref{fig: Fig_5}(a)-(d), and is one of the main results of this paper. %\oa{[The imaginary parts of the term one and two cancel out, but I can't find a way to reduce the equation.]} 
Moreover, in Fig.~\ref{fig: Fig_5}(e), we can plot the shift $\Omega(N,M)$ for every partition $M>0$ as a function of network size.

\begin{table}[tb]
	\renewcommand{\arraystretch}{1.5}
	\centering
	\begin{tabular}{c|c|c||c|c|c}
		\hline
		\ $N=2$ \  & \ \textbf{example} \  & \ \textbf{type} \  &
		\ $N=3$ \  & \ \textbf{example} \  & \ \textbf{type} \
		\\
		\hline
		$\{2,0\}$
		           &
		$\ket{\n\n}$
		           &
		S
		           &
		$\{3,0\}$
		           &
		$\ket{\s\s\s}$
		           &
		S
		\\
		\hline
		$\{1,1\}$
		           &
		$\ket{\s\n}$
		           &
		A
		           &
		$\{2,1\}$
		           &
		$\ket{\s\s\n}$
		           &
		M
		\\
		\hline
		\hline
		$N=4$      & \textbf{example}      & \textbf{type}      &
		$N=5$      & \textbf{example}      & \textbf{type}
		\\
		\hline
		$\{4,0\}$
		           &
		$\ket{\s\s\s\s}$
		           &
		S
		           &
		$\{5,0\}$
		           &
		$\ket{\n\n\n\n\n}$
		           &
		S
		\\
		\hline
		$\{3,1\}$
		           &
		$\ket{\s\n\n\n}$
		           &
		M
		           &
		$\{4,1\}$
		           &
		$\ket{\n\n\s\n\n}$
		           &
		M
		\\
		\hline
		$\{2,2\}$
		           &
		$\ket{\n\s\n\s}$
		           &
		A
		           &
		$\{3,2\}$
		           &
		$\ket{\n\s\n\s\s}$
		           &
		M
		\\
		\hline
		\hline
		$N=6$      & \textbf{example}      & \textbf{type}      &
		$N=7$      & \textbf{example}      & \textbf{type}
		\\
		\hline
		$\{6,0\}$
		           &
		$\ket{\n\n\n\n\n\n}$
		           &
		S
		           &
		$\{7,0\}$
		           &
		$\ket{\s\s\s\s\s\s\s}$
		           &
		S
		\\
		\hline
		$\{5,1\}$
		           &
		$\ket{\s\n\s\s\s\s}$
		           &
		M
		           &
		$\{6,1\}$
		           &
		$\ket{\n\n\n\n\s\n\n}$
		           &
		M
		\\
		\hline
		$\{4,2\}$
		           &
		$\ket{\s\s\n\s\s\n}$
		           &
		M
		           &
		$\{5,2\}$
		           &
		$\ket{\n\s\n\n\s\n\n}$
		           &
		M
		\\
		\hline
		$\{3,3\}$
		           &
		$\ket{\n\s\n\s\n\s}$
		           &
		A
		           &
		$\{4,3\}$
		           &
		$\ket{\s\s\n\n\s\n\n}$
		           &
		M
		\\
		\hline
	\end{tabular}
	\caption{
		Summary of partitions for networks of $N=2$ to $N=7$ equally all-to-all coupled KPOs.  An example of a configuration for each partition  is shown in the bare basis $\mathcal{J}$, along with the type of normal-mode symmetry of the partition, as defined by the coupling matrix [cf.~Sec.~\ref{sec:stability linear}]: symmetric (S), antisymmetric (A), or mixed-symmetry (M). The mixed-symmetry nature of the partition yield a non-reciprocal coupling in their respective EOMs~\eqref{eq: normal mode all-to-all1} and~\eqref{eq: normal mode all-to-all1}. \label{tab: partitions}}
\end{table}

There are two cases that can be understood intuitively: (i) for the partition $\{N,0\}$ (ferromagnetic, $M=N$), we have $x_n=x_m$ (such that $x_a=0$), and the shift is determined by the frequency $\Omega(N,N)=1-N$. Thus, we obtain the non-shifted instability line of the symmetric normal mode; (ii) for the partition $\{n,n\}$ (antiferromagnetic; $N$ even), the non-reciprocal coupling vanishes in Eqs.~\eqref{eq: normal mode all-to-all1} and~\eqref{eq: normal mode all-to-all2}, and reduces to the $N=2$ case solved in Sec.~\ref{sec: 2 id KPO}. Thus, we find $\Omega(N,0)= (2 + N) / 2$, which we draw as the red line in Fig.~\ref{fig: Fig_5}(e). For other partitions, the non-reciprocal coupling remains. This aligns with the fact that these partitions represent states of mixed normal mode symmetry in the network, cf.~Tab.~\ref{tab: partitions}. The frequency shifts of these partitions are irrational, e.g. $\Omega(3,1) = 2+\sqrt{6 \sqrt{3}+9}$, and are larger than those of both the trivial $\{n,n\}$ and $\{N,0\}$ partitions.

The frequency shift $\Omega(N,M)$ in Eq.~\eqref{eq: shift} increases monotonically with $M$ (excluding $M=N$), reflecting the fact that states with a larger imbalance between up and down spins require more energy to stabilize. This behavior stems from the tendency of coupled oscillators to synchronize ($J>0$). When more oscillators are in phase (pointing up or down together), they reinforce each other's motion through the coupling. In contrast, when oscillators are out of phase, they work against each other, requiring a higher parametric drive amplitude $\lambda$ to maintain stability. Hence, the ferromagnetic and antiferromagnetic states are the first to become stable as we increase the parametric drive, while states with mixed up and down configurations require progressively higher drive amplitudes to overcome the effect of the coupling.

Having obtained an analytical expression for all Ising bifurcation lines, we can examine their behavior in the thermodynamic limit $N\to\infty$. Most notably, the frequency difference between consecutive bifurcation lines approaches a constant value, as seen in Fig.~\ref{fig: Fig_5}(f). We have that
\begin{equation}
	\label{eq: shift limit 1}
	\Omega(N, M)-\Omega(N, M+2)\approx\frac{9}{2} +\mathcal{O}(1/N)\,.
\end{equation}
Consequently, in the large-$N$ limit, the bifurcation lines become uniformly spaced, resulting in a highly regular structure in the phase diagram that can be predicted analytically.

Similarly, we analyze the change in $\Omega(N, M)$ when increasing the network size. As shown in Figs.~\ref{fig: Fig_5}(e) and (f), the frequency shift is quasi-linear with a slope of
\begin{equation}
	\label{eq: shift limit 2}
	\Omega(N, M)-\Omega(N+1, M)\approx3+\mathcal{O}(1/N^{4/3})\,.
\end{equation}
In other words, all partition bifurcation lines gain an additional frequency shift of $3J$ when adding a node to the network in the $N\to\infty$ limit.

In particular, we have that the bifurcation line $\lambda_{\{N-1,1\}}$ becomes
\begin{equation}\label{eq: bif line limit}
	\lambda_{\{N-1,1\}}\approx(6N-7)J-2\omega^2+2\omega_0^2 + \mathcal{O}(1/N^{2/3}).
\end{equation}
Together with $\lambda^{(1/2)}_+$ [cf. Eq.~\eqref{eq: bif line bare modes}], this expression localizes the Ising regime, i.e, the region in parameter space where the systems parametrically excited $2^N$ states that can be mapped onto the states of an Ising Hamiltonian.

\section{Conclusion}\label{sec: conclusion}
We have applied secular perturbation theory, to understand the phase diagram of coupled Kerr parametric oscillators. We demonstrated our approach using a network of two coupled KPOs, where we showed how the interplay between parametric driving and coupling leads to state-dependent bifurcation lines that can be understood through perturbative analysis in two limits: when the coupling dominates ($\lambda \ll J$) and when the parametric drive dominates ($\lambda \gg J$). This understanding allows us to predict when the antisymmetric state is stable, as well as when additional mixed-symmetry states emerge.

We extended our analysis to larger networks with all-to-all equal coupling, where we found that the states can be classified by partitions based on the number of spins pointing up or down. Each partition undergoes a period-doubling bifurcation at a frequency that we determine analytically. In the thermodynamic limit, we show that the spacing between bifurcation lines becomes uniform, with a predictable shift as the system size increases.

To successfully utilize the system as a $\frac{1}{2}$-spin Ising network, it is imperative that the solution space consists of exactly $2^N$ distinct configurations, corresponding to all possible combinations of spin orientations. Our analytical framework delineates the precise boundaries of this regime by deriving the bifurcation lines that govern the stability of these configurations. This comprehensive analysis extends beyond previous studies~\cite{DykmanInteraction2018,HeugelIsing2022}, proving that the operating regime exists at finite driving strength and detuning. Our result in Sec.~\ref{sec: N id J}, allows experimentalists to assess the parametric driving required to reach the Ising solution space in their particular implementation depending on the coupling a detuning of the parametric drive. We thus prove that this Ising regime exists and provide a guide how to access it.

Future work will explore the impact of different coupling topologies and the inclusion of disorder in the coupling strengths on the phase diagram and bifurcation structure. Furthermore, the exact error of the difference in the energy landscape of the KPO network and Ising machine remains to be explored.

% \section*{Code Availability}
% Codes to reproduce the results and figures are available on GitHub at Ref.~\cite{ZenodoRepo}.

\begin{acknowledgments}
	We thank the valuable feedback and insightful discussions contributed by Markus Bestler and Gabriel Margiani. We acknowledge funding from the Deutsche Forschungsgemeinschaft (DFG) via project number 449653034 and through SFB1432, as well as the Swiss National Science Foundation (SNSF) through the Sinergia Grant No.~CRSII5\_206008/1.
\end{acknowledgments}

\bibliography{bibliography.bib}

\newpage
\appendix

\section{Relation between the quantum and classical system}
\label{sec: relation Q and C}
\subsection{The classical system}
The Hamiltonian of $N$-coupled Kerr Parametric Oscillators (KPOs) is given by
\begin{multline}\label{SM: Classical Hamiltonian KPO}
	H
	=\sum_i\frac{{p}^2_i}{2 m_i}
	+\frac{m_i}{2}\left(\omega_{i}^2-\lambda_i\cos\left({2 \omega t}\right)\right)x_i^2\\
	+\frac{\alpha_i}{4} {x}^4_i
	-\frac{1}{2}\sum_{j \neq i} J_{i j} x_i x_j,
\end{multline}
where $m_i$ is the mass of the $i^{\rm th}$ oscillator described by the phase space coordinates $x_i$ and $p_i$, $\omega_{i}$ is the bare (angular) frequency, and $\alpha_i$ the Kerr nonlinearity. All oscillators are driven around $\omega \approx \omega_i= \omega_0$ by a parametric pump of strength $\lambda_i$. The network is formed by linearly coupling the KPOs with a coupling strength $J_{i j}$. Commonly, units are chosen such that the masses $m_i$ are unity. The Hamiltonian $H$ yields the $N$ coupled second-order Hamilton's equations of motion \eqref{eq: eom KPO} from the main text. To account for energy dissipation in the system, we introduced linear damping terms with coefficients $\gamma_i$ in the EOMs.

We analyze the system under the assumption that the primary response is at half the frequency of the parametric drive $\omega$. To isolate this response, we rotate the system to that frequency. We do this using the transformation:
\begin{align} \label{sm: van_der_pol}
	x & = u \cos(\omega t) + v \sin(\omega t),                       \\
	p & = - \omega \, u \sin(\omega t) + \omega \, v \cos(\omega t),
\end{align}
where the quadrature $u$ is the new generalized coordinate and the quadrature $v$ the new generalized conjugate momentum. Given that the Poisson bracket $\{x, p\} = 1$, we have that for the new coordinates $\{u, v\} = \frac{1}{\omega}$.
The transformation is a canonical transformation as it can be generated from the type-II generating function
\begin{equation}
	S_2 \left(x , v , t\right) = \omega v x \sec \left(\omega t\right) - \frac{\omega}{2} \left(v^2 + x^2\right) \tan \left(\omega t\right).
\end{equation}
\begin{figure}[t]
	\centering
	\includegraphics[width=\linewidth]{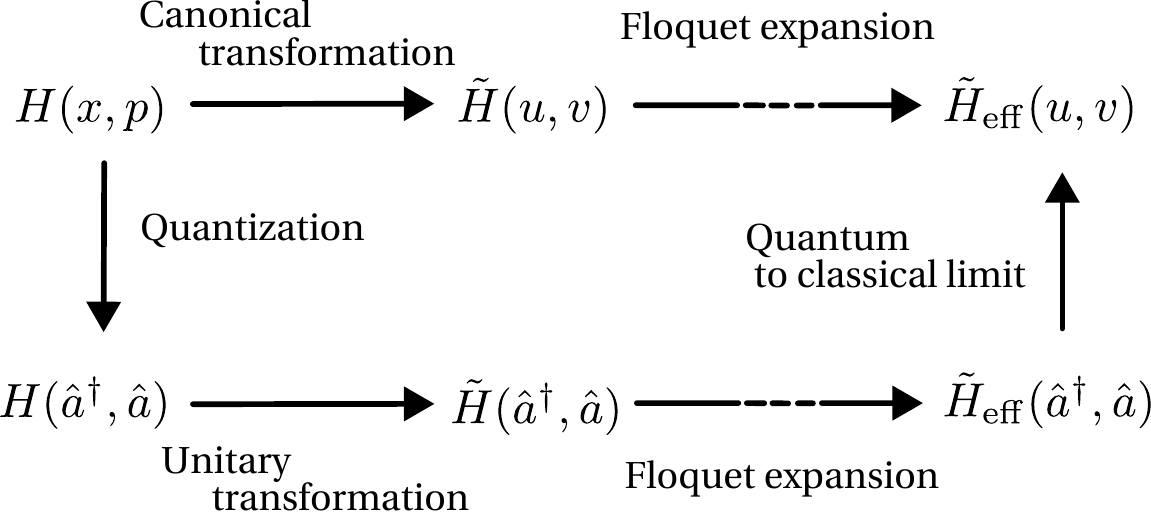}
	\caption{Diagram highlighting the correspondence between the classical Krylov-Bogoliubov and quantum van Vleck Floquet expansions to obtain an effective Hamiltonian for the stroboscopic dynamics.}
	\label{fig: Hamiltonian_averaging}
\end{figure}
Indeed, using that $u=\frac{1}{\omega} \pdv{S_2}{v}$ and $p=\pdv{S_2}{x}$, we can compute that
\begin{align}
	u & = x \sec(\omega t) - v \tan(\omega t)\,,               \\
	p & = \omega v \sec(\omega t) - \omega x \tan(\omega t)\,,
\end{align}
where its solution yields Eq.~\eqref{sm: van_der_pol}. The $1 / \omega$ in the definition of $u=\frac{1}{\omega} \pdv{S_2}{v}$ arises from the fact that the Poisson bracket between the generalized coordinate and the momenta changed from unity to $1 / \omega$. It fixes the dimensions of the variables as both the v and u have units of length.

The Hamiltonian in Eq.~\eqref{SM: Classical Hamiltonian KPO} in the new basis is given by
\begin{widetext}
	\begin{equation} \label{sm: cl rotating Hamiltonian}
		\begin{aligned}
			\tilde{H}(u,v,t) & = H(x,p,t) + \pdv{S_2}{t}                                                                                                                                             \\
			                 & = \frac{3}{32} \alpha (u^2+v^2)^2 + \frac{\lambda}{8}(v^2-u^2) - \frac{1}{4}(u^2+v^2) (\omega^2- \omega_0^2)                                                          \\
			                 & \quad + \frac{1}{4} u v \sin(2 \omega t) (\alpha u^2 + \alpha v^2 - 2 \omega^2 + 2 \omega_0^2)
			+ \frac{1}{8} \cos(2 \omega t) (\alpha u^4 - 2 u^2 (\lambda + \omega^2 - \omega_0^2) - \alpha v^4 - 2 v^2 (\lambda - \omega^2 + \omega_0^2))                                             \\
			                 & \quad + \frac{1}{32} \cos(4 \omega t) (4 \lambda (v^2 - u^2) + \alpha (u^4 - 6 u^2 v^2 + v^4)) + \frac{1}{8} u v \sin(4 \omega t) (\alpha (u^2 - v^2) - 2 \lambda)\,.
		\end{aligned}
	\end{equation}
\end{widetext}
The transformation separates the fast oscillating terms at frequency $2\omega$ and $4\omega$ from the stroboscopic dynamics at $\omega$. A new effective Hamiltonian can be computed by applying a high frequency expansion \cite{burshteinHamiltonian1962}:
\begin{multline} \label{sm: hamiltonian KB}
	H_{\mathrm{eff}}
	=\epsilon [H]_{\mathrm{av}}+\frac{\epsilon^2}{2}\left[\{\check{H}, H\}\right]_{\mathrm{av}}
	\\
	+\frac{\epsilon^3}{3}\left\{\check{H},\left\{\check{H}, H+\frac{[H]_{\mathrm{av}}}{2}\right\}\right\} + \mathcal{O}(\epsilon^4)\,,
\end{multline}
with $\check{H}=\overline{\int \bar{H} \dd{t}}$,  $\bar{H}=H-[H]_{\mathrm{av}}$ and the brackets representing the time average $[\,\cdot\,]_\mathrm{av}\equiv\frac{\omega}{\pi}\int_0^{\pi/\omega} \cdot \, \dd{t}$. Here, the expansion parameter $\epsilon$ corresponds to the one defined in Sec.~\ref{sec: Beyond parametric instability}. Eq.~\eqref{sm: hamiltonian KB} is more often written in terms of Fourier components of the Hamiltonian:
% \begingroup\small
% \medmuskip=0.1mu
% \thinmuskip=0.1mu
% \thickmuskip=0.1mu
\begin{multline} \label{sm: hamiltonian KB fourier}
	H_{\mathrm{eff}}
	=\epsilon H^{(0)}+\epsilon^2\sum_{n\neq0}\frac{\{H_n,H_{-n}\}}{2 n\omega}+\\
	\epsilon^3\left(\sum_{n\neq0}
	\frac{\{H_{-n},\{H_0, H_{n}\}\}}{2(n  \omega)^2}
	+ \sum_{\genfrac{}{}{0pt}{}{m \neq 0}{n}} \frac{\{H_{-m},\{H_{m-n}, H_{n}\}\}}{3n m \omega^2}
	\right),
\end{multline}
% \endgroup
where $H_{n}=\left[ e^{-i n\omega t} H\right]_\mathrm{av}$ is the Fourier component of the Hamiltonian. The effective Hamiltonian in Eq.~\eqref{sm: hamiltonian KB} can be used to study the stroboscopic dynamics of the system.

To first order, we obtain the effective Hamiltonian
\begin{multline} \label{sm: hamiltonian KB first order}
	H_\mathrm{eff}(u,v,t) = \sum_i \frac{3}{32} \alpha (u_i^2+v_i^2)^2 + \frac{\lambda}{8}(v_i^2-u_i^2) \\
	- \frac{1}{4}(u_i^2+v_i^2) (\omega^2- \omega_0^2) + \sum_{j \neq i} \frac{J_{ij}}{2} (u_i u_j + v_j v_i)\,.
\end{multline}
Computing Hamilton's equations of motion for the effective Hamiltonian, we find the slow-flow equations in Eqs.~\eqref{eq: slow-flow equations1}~and~\eqref{eq: slow-flow equations2} in the main text, excluding dissipation. The same expansion can be derived for the full equations of motion in Eq.~\eqref{eq: eom KPO} including dissipation using a near-identity transformation~\cite{holmesSecond1981,seibold2024floquetexpansioncountingpump}.
% So that for the rotated equations of motion $\dot{\mathbf{d}}=\mathbf{F}_{\mathbf{d}}$ with transformation~\eqref{sm: van_der_pol}, we have
% \begingroup\small
% \medmuskip=0.1mu
% \thinmuskip=0.1mu
% \thickmuskip=0.1mu
% \begin{equation}
%   \dot{\vb{D}}_0  =\epsilon\left[-\dot{\vb{D}}_1+\vb{F}_{\vb{d}}\right]+\epsilon^2\left[-\dot{\vb{D}}_2+\vb{G}_{\vb{d}}\right] + \mathcal{O}(\epsilon^3)\,,
% \end{equation}
% \endgroup
% where $\vb{D}_0$ is the lowest order time-dependence. The generating function $\vb{D}_i$ are defined up to second order by $ \dot{\mathbf{D}}_1=\mathbf{F}_{\mathbf{d}}-[\mathbf{F}_{\mathbf{d}}]_\mathrm{av}$ and $\dot{\mathbf{D}}_2=\mathbf{G}_{\mathbf{d}}-[\mathbf{G}_{\mathbf{d}}]_\mathrm{av}$ where
% \begin{equation}
%   \mathbf{G}_{\mathbf{d}} \equiv \mathbf{F}_{\mathbf{d}}^{\prime} \mathbf{D}_1-\mathbf{D}_1^{\prime}\left(-\dot{\mathbf{D}}_1+\mathbf{F}_{\mathbf{d}}\right),
% \end{equation}
% with ${}^\prime$ the Gradient operator for the slow-flow variables $\vb{D}_0$. For the system to remain canonical, we need to impose that $[\,D_1\,]_\mathrm{av}=0$. For more detail see the supplemental material in Ref.~\cite{seibold2024floquetexpansioncountingpump}.

\subsection{The quantum system}

The same Floquet expansion can be performed in the quantum formalism, as highlighted in Fig.~\ref{fig: Hamiltonian_averaging}. Starting from the Hamiltonian in Eq.~\eqref{eq: non-RWA Qham}, one similarly rotates the system to half the frequency of the parametric drive with the unitary transformation $\hat{U}(t) = e^{i \omega t a^\dagger a}$, where $a$ is the annihilation operator counting photon with energy of the drive $\hbar\omega$. The transformed Hamiltonian is given by
\begin{widetext}
	\begin{align}
		\tilde{H}(t) & =
		\tilde{H}_\mathrm{RWA}+e^{-2 i t \omega }\tilde{H}_{-2}+e^{2 i t \omega }\tilde{H}_{2}+e^{-4 i t \omega }\tilde{H}_{-4}+e^{4 i t \omega }\tilde{H}_{4}\,,                                                                                                                                                                                                                                                                                             \\
		             & =-\hbar\Delta\hat{a}^{\dagger } \hat{a}+\frac{\hbar U}{2}(2\hat{a}^{\dagger } \hat{a}+\hat{a}^{\dagger } \hat{a}^{\dagger } \hat{a} \hat{a}) + \frac{\hbar G}{2}(\hat{a}^\dagger \hat{a}^\dagger+\hat{a} \hat{a})                                                                                                                                                                                                                      \\
		             & \quad+e^{-2 i t \omega } \left(\frac{\hbar U}{3} \hat{a} \hat{a} \hat{a}+\frac{\hbar U}{2}\hat{a} \hat{a}-\frac{\hbar\Delta}{2}\hat{a} \hat{a}-\hbar G\hat{a}^\dagger\hat{a}\right)+e^{2 i t \omega } \left(\frac{\hbar U}{3}\hat{a}^{\dagger } \hat{a}^{\dagger } \hat{a}^{\dagger }+\frac{\hbar U}{2}\hat{a}^{\dagger } \hat{a}^{\dagger }-\frac{\hbar\Delta}{2}\hat{a}^\dagger \hat{a}^\dagger-\hbar G\hat{a}^\dagger\hat{a}\right) \\
		             & \quad+e^{-4 i t \omega } \left(\frac{\hbar U}{12}\hat{a} \hat{a} \hat{a} \hat{a}-\frac{\hbar G}{2}\hat{a}\hat{a}\right)+e^{4 i t \omega } \left(\frac{\hbar U }{12}\hat{a}^{\dagger } \hat{a}^{\dagger } \hat{a}^{\dagger } \hat{a}^{\dagger }-\frac{\hbar G}{2}\hat{a}^\dagger\hat{a}^\dagger\right)\,,
	\end{align}
\end{widetext}
with $\Delta=\frac{(\omega^2-\omega_0^2)}{2\omega}$, $U=\frac{3 \alpha  \hbar }{4 \omega^2}$ and $G=\frac{\lambda}{4\omega}$.
Similarly, to the classical rotated Hamiltonian in Eq.~\eqref{sm: cl rotating Hamiltonian}, the processes at $\omega$ are time-independent with only additional terms acting at short time scales $1/2\omega$ and $1/4\omega$. Having isolated the stroboscopic dynamics, one typically applies the van Vleck degenerate perturbation theory~\cite{EckardtHighFrequency2015}. The effective Hamiltonian is then given by Eqs.~\eqref{sm: hamiltonian KB} or~\eqref{sm: hamiltonian KB fourier} with replacing the Poisson brackets with the associated quantum commutator $\{\cdot,\cdot\}\to \frac{1}{i\hbar} [\cdot,\cdot]$. We can take the quantum-to-classical limit by taking the mean field approximation $\hat{a}\approx\expval{\hat{a}}$ and $\hbar\to 0$. Additionally, splitting the coherent field into its imaginary and real part (quadratures) $\expval{\hat{a}}=  \sqrt{\frac{ \omega }{ 2 \hbar}}(u-i v)$, we find the same effective Hamiltonian as the classical system.
%  \oa{[Check if the $1/\omega$ is correct.]}

\section{Ising machine}
\label{sec: Ising machine}

In the limit where the parametric drive $\lambda$ is much larger than the coupling $J$, the first order classical effective Hamiltonian of the KPO network~\eqref{sm: hamiltonian KB first order} can be mapped to an Ising Hamiltonian. To see this, we write the Hamiltonian in terms of the action-angle variables of the harmonic oscillator:
\begin{equation}
	u_i = A_i \cos(\phi_i) \quad \text{and} \quad v_i = A_i \sin(\phi_i),
\end{equation}
where $A_i>0$ and $0 \leq \phi_i \leq 2\pi$. Using a type I generating function $F_1(u_i, \phi_i) = \frac{1}{2} u_i^2 \cot(\phi_i)$, we obtain the non-interacting Hamiltonian in terms of these action variables:
% \begingroup\small
% \medmuskip=0.1mu
% \thinmuskip=0.1mu
% \thickmuskip=0.1mu
\begin{equation}
	H_0 =\sum_i \frac{3}{32} \alpha A_i^4 - \frac{1}{8} \left( 2 \omega^2 - 2 \omega_0^2 + \lambda \cos (2 \phi_i) \right) A_i^2.
\end{equation}
% \endgroup
The interaction Hamiltonian is given by:
\begingroup\small
\medmuskip=0.1mu
\thinmuskip=0.1mu
\thickmuskip=0.1mu
\begin{equation}
	H_I = \sum_{j \neq i} \frac{J_{ij}}{2} A_i A_j \left( \cos(\phi_i) \cos(\phi_j) + \sin(\phi_i) \sin(\phi_j) \right).
\end{equation}
\endgroup
We can deduce from the EOMs~\eqref{eq: eom KPO} that when we have no dissipation, $v_i = A_i \sin(\phi_i) \rightarrow 0$, such that $\phi_i = 0$ or $\pi$. Therefore, defining $\sigma_i = \cos(\phi_i)$, we can write the Hamiltonian as:
\begin{multline} \label{sm: Ising Hamiltonian}
	H_{\text{Ising}} = \sum_i \left( \frac{3}{32} \alpha A_i^4 - \frac{1}{8} \left( 2 \omega^2 - 2 \omega_0^2 + \lambda \right) A_i^2 \right)\\
	+ \sum_{j \neq i} \frac{J_{ij}}{2} A_i A_j \sigma_i \sigma_j.
\end{multline}
Given that the network is in the Ising regime [cf. Sec.~\ref{sec: N id J}], the Hamiltonian reduces to the Ising Hamiltonian up to a rescaling of the ground state energy when the amplitude of the oscillators are equal. Note that using a perturbation in the elements of the coupling matrix, i.e., in the limit $J_{ij} \ll 1$, we can compute that
\begingroup\small
\medmuskip=0.1mu
\thinmuskip=0.1mu
\thickmuskip=0.1mu
\begin{equation}
	\resizebox{.87\hsize}{!}{$\displaystyle A_i \approx \sqrt{\frac{2(2 \omega^2 - 2 \omega_0^2 + \lambda)}{3 \alpha}} + \frac{\sqrt{\frac{2}{3}} J}{\sqrt{\alpha} \sqrt{\lambda + 2 \omega^2 - 2 \omega_0^2}} + \mathcal{O}(J^2)$}\,.
\end{equation}
\endgroup
Hence, given $ \sqrt{\lambda + 2 \omega^2 - 2 \omega_0^2} \gg 1$ or $J_{ij} \ll 1$ the phases states of the oscillators effectively act as spins in the Ising Hamiltonian.

\begin{figure}
	\centering
	\includegraphics[width=\linewidth]{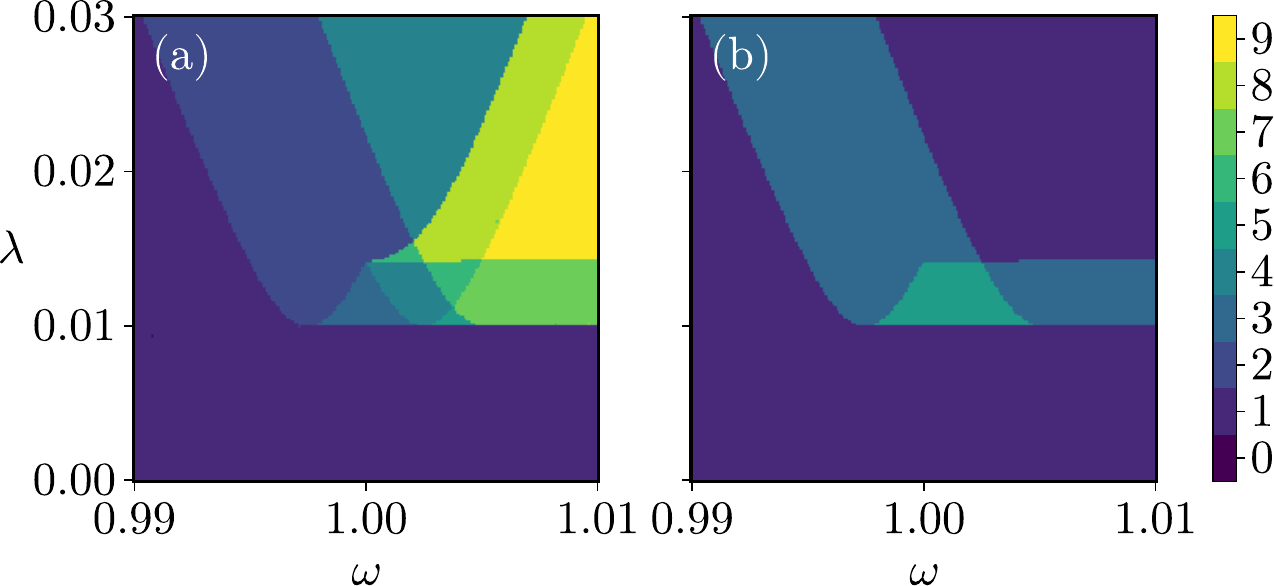}
	\caption{Two damped coupled KPOs. (a) Phase diagram visualized by plotting the number of stable states as a function of the parametric pump amplitude $\lambda$ and driving frequency $\omega$. (b) The stability of the symmetric normal mode $x_s$, cf. Eq.~\eqref{SM: shifted mathieu lobe damping}, perturbed by the nonlinear interaction of a fixed antisymmetric phase state.
	}
	\label{fig: Fig_A1}
\end{figure}

\section{Two coupled parametric resonators}
\label{sec: extr info 2}
\subsection{No damping}
We provide here additional details for the analysis discussed in Sec.~\ref{sec: 2 id KPO} of the main text, consisting of two identical KPOs with bilinear coupling coefficient $J$.

All calculations are performed within the rotating wave approximation (first order KB). As such, for reference, we provide the slow-flow equations in both the bare mode and normal mode bases. In the bare mode basis, we have:
\begin{align}\label{SM: averaged eom bare basis}
	\dot{v}_1 & =\frac{J u_2}{2 \omega }+u_1\left(\frac{3 \alpha  u_1^2}{8 \omega }-\frac{ \lambda}{4 \omega } -\frac{ \omega ^2- \omega_1^2}{2 \omega }\right), \\
	\dot{v}_2 & =\frac{J u_1}{2 \omega }+u_2 \left(\frac{3 \alpha  u_2^2}{8 \omega }-\frac{\lambda}{4 \omega } -\frac{ \omega ^2- \omega_2^2}{2 \omega }\right).
\end{align}
While in the normal mode basis, we obtain:
\begin{align}\label{SM: averaged eom normal basis}
	\dot{v}_s & =u_s \left(\frac{3 \alpha u_s^2}{32 \omega }-\frac{\lambda}{4 \omega }-\frac{ \omega ^2- \omega_s^2}{ 2\omega }+\frac{9 \alpha  u_a^2}{32 \omega }\right), \\
	\dot{v}_a & =u_a \left(\frac{3 \alpha u_a^2}{32 \omega }-\frac{\lambda}{4 \omega }-\frac{ \omega ^2- \omega_a^2}{2 \omega }+\frac{9 \alpha  u_s^2}{32 \omega }\right).
\end{align}

The bifurcation lines observed in Fig.~\ref{fig: Fig_2} can be derived analytically by analyzing the stability of the system's fixed points. This analysis involves computing the Jacobian of Eqs.~\eqref{SM: averaged eom bare basis} and \eqref{SM: averaged eom normal basis}, evaluating it at the stationary solutions, and determining stability through the real parts of its eigenvalues. This procedure yields the bifurcation lines presented in Sec.~\ref{sec: 2 id KPO} of the main text.

\subsection{With damping}
For simplicity, we have performed our analysis in the no dissipation limit $\gamma_i=\gamma=0$. Instead, for a finite damping coefficient, the stationary state $x(t)=A\cos(\omega t + \theta)$ of an uncoupled parametric oscillator gains a phase, computed to be
\begin{equation}
	\theta = \arctan(\frac{\sqrt{\lambda ^2-4 \gamma ^2 \omega ^2}-\lambda }{2 \gamma  \omega }),
\end{equation}
or
\begin{equation}
	\theta = \arctan(\frac{\sqrt{\lambda ^2-4 \gamma ^2 \omega ^2}-\lambda }{2 \gamma  \omega }) + \pi.
\end{equation}
If we dress the symmetric state with a fixed stationary state of the antisymmetric state, we find the dressed EOM:
\begingroup
\medmuskip=0.1mu
\thinmuskip=0.1mu
\thickmuskip=0.1mu
\begin{multline}\label{SM: shifted mathieu lobe damping}
	\ddot{x}_s
	+ \gamma \dot{x}_s= \\
	-\left[\omega_s^2-\lambda\cos(2\omega t)+\frac{3\alpha}{2}A_a^2(1+\cos(2\omega t+2\theta_a))\right]x_s .
\end{multline}
\endgroup
Analytically computing the bifurcation lines of Eq.~\eqref{SM: shifted mathieu lobe damping} becomes impractical as the fixed normal mode drives the system with a different phase than that of the parametric drive. Instead, in Sec.~\ref{sec: secular perturbation}, we compute the bifurcation lines analytically with damping using secular perturbation theory. Here, we can still compute the stability lines numerically, which results in the dressed phase diagram in Fig.~\ref{fig: Fig_A1}(b). The lines exactly correspond to the ones found in the full phase diagram in Fig.~\ref{fig: Fig_A1}(a).

% %
% \begin{figure}
%   \centering
%   \includegraphics[width=\linewidth]{figures/suppmat/inhomo_two_parametrons/Fig4.pdf}
%   \caption{Two detuned coupled KPOs. (a) Phase diagram visualized by plotting the number of stable states as a function of the parametric pump amplitude $\lambda$ and driving frequency $\omega$ where $\delta\approx J$ [cf. purple in (c) and (d)]. (b) Phase diagram in the regime $\delta\gg J$ [cf. orange in (c) and (d)] (c) Eigenfrequencies of two detuned KPO's (purple and orange lines) in function of the detuning parameter $\delta=(\omega_2-\omega_1)/2$ between the bare frequencies, showing avoided crossing. The gray dashed lines show the uncoupled case. (d) Heatmap of the rotational angle $\theta(\delta, J)$ associated with the normal mode basis transformation. The angle resembles the division of two extremal regimes: $\theta=0\degree$ (red) where the system is effectively decoupled $(\bar{J}=0)$ and $\theta=45\degree$ (blue) where the system is effectively tuned $(\bar{\delta}=0)$.}
%   \label{fig: Fig_4}
% \end{figure}  
%

\section{Secular perturbation: the case of two parametrons}
\label{sec: secular perturbation}
In this section, we employ the Poincaré–Lindstedt secular perturbation method to rederive the bifurcation lines in Eq.~\eqref{eq: bif lines symmetric state}. This technique systematically eliminates secular terms (that grow unbounded in time) by introducing a small parameter expansion of both the solution and time variables. This approach guarantees uniformly valid approximations over long time scales, making it particularly suitable for analyzing the long-term dynamics of driven oscillators.

We start from the equations of motion in the normal mode basis:
\begingroup
\medmuskip=0.1mu
\thinmuskip=0.1mu
\thickmuskip=0.1mu
\begin{align}
	 & \ddot{x}_s +\left[\omega_s^2 - \lambda \cos(2\omega t)\right]x_s +\frac{\alpha}{4}( x_s^3+3x_a^2x_s)+\gamma \dot{x}_s=0\,,   \\
	 & \ddot{x}_a +\left[\omega_a^2 - \lambda \cos(2\omega t)\right]x_a + \frac{\alpha}{4}( x_a^3+ 3x_s^2x_a)+\gamma \dot{x}_a=0\,.
\end{align}
\endgroup
We are interested in the system's response at the frequency $\omega$. To facilitate this, we rewrite the ODEs in terms of dimensionless time $\tau=\omega t -\phi(\epsilon)$, where $\phi(\epsilon)$ is a phase correction that will be determined later. This yields
\begingroup
\medmuskip=0.1mu
\thinmuskip=0.1mu
\thickmuskip=0.1mu
\begin{align} \label{sm: PL setup 1}
	\ddot{x}_s +x_s & =\left[\Delta_s  + \frac{\lambda}{\omega^2}\cos(2\tau+ 2\phi)- \frac{\alpha}{4\omega^2}( x_s^2+3 x_a^2)\right]x_s -\frac{\gamma}{\omega}\dot{x_s}\,,   \\
	\ddot{x}_a +x_a & = \left[\Delta_a + \frac{\lambda}{\omega^2} \cos(2\tau+ 2\phi) - \frac{\alpha}{4\omega^2}( x_a^2+3 x_s^2)\right]x_a -\frac{\gamma}{\omega}\dot{x_a}\,,
	\label{sm: PL setup 2}
\end{align}
\endgroup
with $\Delta_s= \frac{\omega^2-\omega_s^2}{\omega^2}$ and $\Delta_a= \frac{\omega^2-\omega_a^2}{\omega^2}$. We assume $\epsilon$ to be a small parameter, $0<\epsilon\ll1$, such that we can expand both the displacement $x$ and the phase $\phi$:
\begin{align}
	x_s(\tau)      & =\epsilon x_{s,1}+\epsilon^2x_{s,2}(\tau) \,,\qquad           \\
	x_a(\tau)      & =x_{a,0}+\epsilon x_{a,1}(\tau) +\epsilon^2 x_{a,2}(\tau) \,, \\
	\phi(\epsilon) & =\phi_0+\epsilon \phi_1+\epsilon^2 \phi_2+\ldots \, .
\end{align}
Here, we assume that $x_s(\tau)$ does not have a zero-order contribution, i.e., $x_{s,0}=0$. Additionally, we assume the right-hand side of Eqs.~\eqref{sm: PL setup 1}~and~\eqref{sm: PL setup 2} to be of order $\epsilon$. These assumptions correspond to the same bounds imposed when implying Krylov-Bogoliubov expansion [cf. Eqs.~\eqref{eq: slow-flow equations1}~and~Eqs.~\eqref{eq: slow-flow equations2}].

The zeroth order in $\epsilon$ reads
\begin{align}
	\ddot{x}_{a,0} + x_{a,0} =0\,,
\end{align}
which is easily solved with $x_{a,0}=A_a\cos(\tau)+B_a\sin(\tau)$. The next order reads
\begingroup
\medmuskip=0.1mu
\thinmuskip=0.1mu
\thickmuskip=0.1mu
\begin{align}
	\ddot{x}_{s,1} + x_{s,1} & = 0\,,                                                                                                                                            \\
	\ddot{x}_{a,1} + x_{a,1} & =\left[\Delta_a+\frac{\lambda}{\omega^2} \cos (2 (\tau+\phi_0))x_{a,0}\right]-\frac{\alpha }{4\omega^2} x_{a,0}^3-\frac{\gamma}{\omega}x_{a,0}\,,
\end{align}
\endgroup
Hence, we have $x_{s,0}=A_s\cos(\tau)+B_s\sin(\tau)$ with the same boundary conditions for the symmetric mode. Central to secular perturbation theory, we need to ensure that resonant terms (secular terms) do not grow unbounded. From this, we obtain the secular conditions:
\begin{widetext}
	% \begingroup
	% \medmuskip=0.1mu
	% \thinmuskip=0.1mu
	% \thickmuskip=0.1mu
	\begin{align}
		\left(\Delta_a + \frac{\lambda}{2\omega^2} \cos (2 \phi_0)-\frac{3\alpha}{16\omega^2} (A_a^2+B_a^2)\right) A_a-\left(\frac{\gamma}{\omega} +\frac{\lambda}{2\omega^2} \sin (2 \phi_0)\right)B_a & =0 \,, \\
		\left(\Delta_a + \frac{\lambda}{2\omega^2} \cos (2 \phi_0)-\frac{3\alpha}{16\omega^2} (A_a^2+B_a^2)\right) B_a-\left( \frac{\lambda}{2\omega^2} \sin (2 \phi_0)-\frac{\gamma}{\omega}\right)A_a & =0\,.
	\end{align}
	% \endgroup
\end{widetext}
When disregarding the coupling to the symmetric mode, these are the same conditions as the averaged Eq.~\eqref{SM: averaged eom normal basis} for the steady-state limit. We have the freedom to choose $\phi_0=0$, such that
\begingroup
\medmuskip=0.1mu
\thinmuskip=0.1mu
\thickmuskip=0.1mu
\begin{equation}
	A_a^2=\frac{4 \left(\left(\sqrt{\lambda ^2-4 \gamma ^2 \omega ^2}+\lambda \right) (\lambda +2 \Delta_a\omega^2)-4 \gamma ^2 \omega ^2\right)}{3 \alpha  \lambda }\,,
\end{equation}
\endgroup
and
\begingroup
\medmuskip=0.1mu
\thinmuskip=0.1mu
\thickmuskip=0.1mu
\begin{equation}
	B_a^2=\frac{4 \left(\left(\sqrt{\lambda ^2-4 \gamma ^2 \omega ^2}-\lambda \right) \left(\lambda -2 \Delta_a\omega^2\right)+4 \gamma ^2 \omega ^2\right)}{3 \alpha  \lambda }\,.
\end{equation}
\endgroup
% Hence, the first-order displacement of the antisymmetric mode is solved to be
% \begin{align}
% x_{a,1}= B_a \cos (\tau)+ \frac{G}{4} \cos (\tau)+\frac{G}{8} \cos (3 \tau)\,,
% \end{align}
% with $G=\frac{1}{16}(\alpha A_a^3 -8 \lambda A_a)$.
The next order EOM for the symmetric mode reads
\begingroup
\medmuskip=0.1mu
\thinmuskip=0.1mu
\thickmuskip=0.1mu
\begin{multline}
	\ddot{x}_{s,2} + x_{s,2} =\\
	\frac{\lambda}{\omega^2} \cos (2 \tau)x_{s,0}-\frac{\alpha }{4\omega^2} x_{s,0}^3+\Delta_s x_{s,0}-\frac{3\alpha}{4\omega^2}x_{a,0}^2x_{s,0}\,.
\end{multline}
\endgroup

The secular condition for the symmetric mode is therefore
The next order EOM for the symmetric mode reads
\begingroup
\medmuskip=0.1mu
\thinmuskip=0.1mu
\thickmuskip=0.1mu
\begin{align}
	 & \resizebox{.83\hsize}{!}{$\displaystyle
	A_{s} \left(-\frac{3\alpha(3 A_{a}^2+ B_{a}^2)}{16\omega^2} + \Delta_s+ \frac{\lambda}{2\omega^2} \right)- \left(\frac{\gamma}{\omega} +\frac{3 A_{a}B_{a}}{8\omega^2}\right)B_s=0$}\,, \\
	 & \resizebox{.83\hsize}{!}{$\displaystyle
			B_{s} \left(-\frac{3\alpha(3 A_{a}^2+ B_{a}^2)}{16\omega^2} + \Delta_s+ \frac{\lambda}{2\omega^2} \right)- \left(\frac{3 A_{a}B_{a}}{8\omega^2}-\frac{\gamma}{\omega} \right)A_s=0$}\,.
\end{align}
\endgroup
If we set $\gamma=0$, we obtain exactly the averaged equation one obtains by applying the first order KB expansion to Eq.~\eqref{eq: mod_eom_normal_mode_2_sym}. Hence, if we substitute the solution for $A_a$ into the secular condition, we obtain the bifurcation lines in Eq.~\eqref{eq: bif lines symmetric state}. When considering the impact of dissipation, we find that it affects the bifurcation lines similarly to how it impacts a single KPO, i.e.,
\begingroup
\medmuskip=0.1mu
\thinmuskip=0.1mu
\thickmuskip=0.1mu
\begin{equation}\label{SM: bif lines symmetric state}
	\lambda^{(s)}_\mathrm{1}
	=\sqrt{4\gamma^2\omega^2+4J^2}
	\qq{and}
	\lambda^{(s)}_\mathrm{2}=\sqrt{4\gamma^2\omega^2+(4 J-2 \omega ^2+2 \omega_0^2)^2}\, .
\end{equation}
\endgroup

\end{document}